%% file: main.tex
\documentclass[acmsmall,acmauthoryear,screen]{acmart}
\AtBeginDocument{%
  }

\setcopyright{acmlicensed}
\copyrightyear{2025}
\acmYear{2025}
\acmDOI{XXXXXXX.XXXXXXX}

\acmISBN{978-1-4503-XXXX-X/18/06}

\usepackage{amsmath,amsfonts}
\usepackage{amssymb}
 \usepackage{algorithmic}
\usepackage{algorithm}
\usepackage{array}
\usepackage[caption=false,font=normalsize,labelfont=sf,textfont=sf]{subfig}
\usepackage{textcomp}
\usepackage{stfloats}
\usepackage{url}
\usepackage{verbatim}
\usepackage{graphicx}

\usepackage{pifont}
\usepackage[ruled,vlined,algo2e]{algorithm2e}
\usepackage[justification=centering]{caption}
\usepackage{enumitem}
\usepackage{subcaption}
\usepackage{booktabs}
\usepackage{makecell}
\usepackage{tcolorbox}
\usepackage{soul}
\usepackage{ragged2e}
\usepackage{threeparttable}
\usepackage{balance}
\usepackage{multirow}
\usepackage{xcolor}
\newcommand{\appname}{{\sc LLM4SZZ}\xspace}
\newcommand{\appnamebold}{{\sc \textbf{LLM4SZZ} \xspace}}
\newcommand{\code}[1]{\texttt{#1}}
\newcommand{\find}[1]{
\begin{tcolorbox}[leftrule=0.5mm,rightrule=0.5mm, toprule=0.5mm,bottomrule=0.5mm,left=2pt,right=2pt,top=2pt,bottom=2pt]%
\em #1
\end{tcolorbox}
}

\newenvironment{coloredtext}[0]{\color{black}}{\ignorespacesafterend}
\begin{document}

\title{LLM4SZZ: Enhancing SZZ Algorithm with Context-Enhanced Assessment on Large Language Models}

\author{Lingxiao Tang}
\authornote{Also with Hangzhou High-Tech Zone (Binjiang) Institute of Blockchain and Data Security}
\email{12421037@zju.edu.cn}
\orcid{0009-0003-7406-7961}
\affiliation{%
  \
  \institution{The State Key Laboratory
 of Blockchain and Data Security, Zhejiang University}
  \city{Hangzhou}
  \state{Zhejiang}
  \country{China}
}

\author{Jiakun Liu}
\email{jkliu@smu.edu.sg}
\orcid{0000-0002-7273-6709}
\affiliation{%
  \institution{Singapore Management University}
  \city{Singapore}
  \country{Singapore}}

\author{Zhongxin Liu}
\email{liu_zx@zju.edu.cn}
\orcid{0000-0002-1981-1626}
\affiliation{%
  \institution{The State Key Laboratory
 of Blockchain and Data Security, Zhejiang University}
  \city{Hangzhou}
  \state{Zhejiang}
  \country{China}
}

\author{Xiaohu Yang}
\email{yangxh@zju.edu.cn}
\orcid{0000-0003-4111-4189}
\affiliation{%
  \institution{The State Key Laboratory
 of Blockchain and Data Security, Zhejiang University}
  \city{Hangzhou}
  \state{Zhejiang}
  \country{China}
}

\author{Lingfeng Bao}
\authornote{Corresponding author}
\authornotemark[1]
\email{lingfengbao@zju.edu.cn}
\orcid{0000-0003-1846-0921}
\affiliation{%
  \institution{The State Key Laboratory
 of Blockchain and Data Security, Zhejiang University}
  \city{Hangzhou}
  \state{Zhejiang}
  \country{China}
}

\renewcommand{\shortauthors}{Tang et al.}

\begin{abstract}
The SZZ algorithm is the dominant technique for identifying bug-inducing commits and serves as a foundation for many software engineering studies, such as bug prediction and static code analysis, thereby enhancing software quality and facilitating better maintenance practices.
Researchers have proposed many variants to enhance the SZZ algorithm's performance since its introduction. 
The majority of them rely on static techniques or heuristic assumptions, making them easy to implement,
but their performance improvements are often limited.
Recently, a deep learning-based SZZ algorithm has been introduced to enhance the original SZZ algorithm. However, it requires complex preprocessing and is restricted to a single programming language. Additionally, while it enhances precision, it sacrifices recall.  
Furthermore, most of variants overlook crucial information, such as commit messages and patch context, and are limited to bug-fixing commits involving deleted lines.

The emergence of large language models (LLMs) offers an opportunity to address these drawbacks. In this study, we investigate the strengths and limitations of LLMs and propose \appname, which employs two approaches (i.e., rank-based identification and context-enhanced identification) to handle different types of bug-fixing commits. We determine which approach to adopt based on the LLM’s ability to comprehend the bug and identify whether the bug is present in a commit. The context-enhanced identification provides the LLM with more context and requires it to find the bug-inducing commit among a set of candidate commits. In rank-based identification, we ask the LLM to select buggy statements from the bug-fixing commit and rank them based on their relevance to the root cause.  Experimental results show that \appname outperforms all baselines across three datasets, improving F1-score by 6.9\% to 16.0\% without significantly sacrificing recall. Additionally, \appname can identify many bug-inducing commits that the baselines fail to detect, accounting for 7.8\%, 7.4\% and 2.5\% of the total bug-inducing commits across three datasets, respectively.

\end{abstract}

\ccsdesc[500]{Software and its engineering~Software maintenance tools}

\keywords{SZZ Algorithm, large language model }

\received{31 October 2024}
\received[revised]{27 February 2025}
\received[accepted]{31 Mar 2025}

\maketitle

\section{Introduction} \label{sec:introduction}
\input{tex/introduction}

\section{Background} \label{sec:background}
\input{tex/background}

\section{Approach} \label{sec:approach}
\input{tex/Approach}

\section{Experiment setup} \label{sec:Experiment setup}
\input{tex/experiment_set_up}

\section{Experiment results} \label{sec:Experiment results}

\input{tex/experiments}

\section{Discussion}
\input{tex/discussion}

\section{Related work}
\input{tex/related_work}

\section{Conclusion and future work}
\input{tex/conclusion_future_work}

\section{Acknowledgement}
This research/project is supported by the National Science Foundation of China (No.62372398 and No.72342025) and the Zhejiang Pioneer (Jianbing) Project (2025C01198(SD2)), and funded by ZJU-China Unicom Digital Security Joint Laboratory.

\section*{Data Availability}
The replication package, which includes the source code, datasets, and LLMs' output, can be found at \url{https://doi.org/10.6084/m9.figshare.27418236.v1}.

\bibliographystyle{ACM-Reference-Format}
\bibliography{refs}

\end{document}

%% file: tex/introduction.tex
Having been proposed in 2005, the SZZ algorithm~\citep{sliwerski2005changes} and its variants have been widely used in finding bug-inducing commits from bug-fixing commits. 
The original SZZ algorithm assumes that the deleted lines in the bug-fixing commit cause the bug. It first locates the deleted lines in the bug-fixing commit. Then, it uses the \texttt{annotate} command from the version control system to trace back the commits that most recently added or modified these lines. Finally, it marks the identified commits as bug-inducing commits. Many downstream tasks can be performed based on bug-inducing commits, such as analyzing why the bugs occur~\cite{bavota2015four,aman2019empirical}, predicting defects~\cite{hata2012bug,yan2020just,fan2019impact}, and measuring the factors that influence software quality~\cite{tufano2017empirical,chen2019extracting}.

Although the SZZ algorithm has achieved great success, it still suffers from low precision. %
Consequently, many variants have been proposed~\cite{kim2006automatic,da2016framework,davies2014comparing,neto2018impact,tang2023neural} to address this problem. Some methods~\cite{kim2006automatic,da2016framework,neto2018impact} attempted %
to improve precision by removing noise in bug-fixing commits using static analysis. Noise refers to changes that do not influence the program's behavior, such as blank lines, comments, or refactoring operations. These irrelevant changes are unrelated to the bug, and tracing them back can lead to false positives in the output.
Other methods~\cite{davies2014comparing} try to improve precision by treating the commits identified by the original SZZ algorithm %
as candidates and selecting the final bug-inducing commit from them. They choose the final bug-inducing commit by considering factors such as commit dates or the number of changed lines. %
To further improve precision, Tang et al.~\cite{tang2023neural} introduced a deep learning method that embeds changed lines based on their semantic meanings and relationships, training a ranking model to identify the deleted lines most likely to cause the bug. However, this approach significantly sacrifices recall.

\begin{coloredtext}
Although previous studies have made some advancements, several limitations still exist. 
\textbf{Limitation 1}: These methods overlook the commit message of the bug-fixing commit. Typically, the commit message contains essential information on why the changes were made~\cite{mockus2000identifying,yan2016automatically} and many of these messages describe how the bug occurs and how the commit fixes it.
This information is vital for understanding the commit and accurately locating buggy statements. 
\textbf{Limitation 2}: These methods assume that only deleted lines cause bugs~\cite{sliwerski2005changes,kim2006automatic,tang2023neural}, making them inapplicable to bug-fixing commits that contain only added lines.
\textbf{Limitation 3}: These methods focus solely on changed lines, ignoring the context of the entire patch. Previous studies have shown that the context, including unmodified lines near the changes, can provide crucial information for the model to understand the code~\cite{chen2019sequencer, xia2023revisiting}.
Sometimes, it might be the unmodified lines themselves that lead to the bug, rather than the changed lines~\cite{rosa2021evaluating}.
\textbf{Limitation 4}: Methods that select the final bug-inducing commit from a set of candidates often rely on heuristic assumptions~\cite{davies2014comparing}, such as commit dates or the number of changed lines. These assumptions may not work in all scenarios~\cite{bao2022v}. %
Ideally, we should determine the final bug-inducing commit based on the root cause of the bug and the content of the candidate commit.

The emergence of Large Language Models (LLMs) presents an opportunity to address the aforementioned limitations. Previous studies indicate that LLMs can effectively understand code changes and commit messages~\cite{li2024understanding,xue2024automated}. One fundamental improvement is to utilize the LLM to analyze the root cause of the bug and identify buggy statements based on code changes and the commit message. This process leverages the commit message, addressing limitation 1. When identifying buggy statements, the LLM can detect not only deleted lines but also unchanged lines, addressing limitation 2. The enhanced approach then traces these buggy statements to obtain a set of candidate commits and requires the LLM to select the bug-inducing commits from this set, addressing limitation 4. Furthermore, we can provide the LLM with more context which solves limitation 3. At first glance, this simple approach seems to address all the problems.
However, several challenges remain in this simple method. 
\textbf{Challenge 1}: LLMs struggle with complex bug-fixing commits that involve numerous changes across multiple files and functions. These commits often contain significant noise unrelated to the bug fix, undermining LLM's performance. \textbf{Challenge 2}: We need to provide more information to help the LLM determine whether the bug exists. The root cause of the bug and the content of the commit are often insufficient (see Section~\ref{sec:Context-enhanced-assessment} and Section~\ref{rq2}).
\textbf{Challenge 3}: When asking the LLM to determine whether a commit contains a bug, we must carefully consider the context provided. An overly long context can degrade performance~\cite{li2023loogle}, while a too-short context may omit crucial information necessary for the LLM to understand the code~\cite{chen2019sequencer,xia2023revisiting}. 
\textbf{Challenge 4}: Many types of bugs remain beyond the LLM's understanding~\cite{parasaram2024fact,bouzenia2024repairagent}, making it difficult for LLMs to ascertain their presence in a commit. Treating these bugs the same way as those the LLM can comprehend will adversely affect overall performance. For instance, if we determine that the LLM can understand the bug and identify its presence in a commit, we can use it to select the final bug-inducing commit from a set of candidates; otherwise, we cannot. Further details will be discussed in Section \ref{rq2}.  Therefore, a better approach is needed to solve those challenges to fully leverage the potential of LLMs.

In this paper, we propose an LLM-based approach called \appname. In the preparation phase, we summarize the root cause of the bug and filter out irrelevant files based on the patch content and the commit message. This step helps eliminate noise when handling large bug-fixing commits, addressing challenge 1. Next, we assess the LLM's ability to understand the bug and the ability to determine whether it exists in the commit. 
Instead of directly asking the LLM to determine whether a commit contains a bug, we employed a more complex strategy, consisting of several parts. This approach is taken because we find that direct judgments are ineffective, see Section~\ref{rq2}.
To evaluate this ability, we first provide the LLM with expanded context and require it to generate a hint indicating whether the bug is present, addressing challenge 2. Before asking the LLM to determine whether the bug exists, we refine the context to address challenge 3.
We then present the LLM with the root cause of the bug, the hint, and refined contexts for two versions of the program: one buggy and one correct. If the LLM can accurately distinguish between the two versions, we consider it capable; otherwise, it is not. 
Based on this ability assessment, we developed two approaches: context-enhanced identification and rank-based identification, which resolves challenge 4. In context-enhanced identification, we provide the LLM with more context and require it to select the bug-inducing commit from a set of candidates. In rank-based identification, we follow the methodology outlined in the previous study~\cite{tang2023neural}, asking the LLM to identify buggy statements from the bug-fixing commit and rank them based on their relevance to the root cause. 
\end{coloredtext}
To evaluate our method, we use three high-quality, developer-annotated datasets, ensuring their accuracy. We assess our proposed method by answering the following questions:

\noindent\textbf{RQ1: How effective is \appnamebold in identifying bug-inducing commits from bug-fixing
commits compared to baselines?}

In this RQ, we compare \appname with all baselines across three datasets to determine whether \appname can outperform the baselines in identifying bug-inducing commits. The experimental results demonstrate that \appname surpasses all other baselines in F1-score, with a notable improvement ranging from 6.9\% to 16.0\%. Furthermore, \appname enhances both precision and F1-score without significantly sacrificing recall.

\noindent\textbf{RQ2: How effective are the key components of \appnamebold?}

We also conduct an ablation experiment to ensure that all key components of \appname, namely the context-enhanced assessment, the context-enhanced identification, and the rank-based identification, contribute to its performance. Additionally, we demonstrate that utilizing LLMs directly on the SZZ algorithm cannot yield satisfactory results.

\noindent\textbf{RQ3: How effective is \appnamebold if we apply it on other open-source large language models?}

In this RQ, we aim to examine whether the core ideas of \appname can be applied to other open-source large language models. We implement \appname using llama3-8b and llama3-70b. The experimental results show that \appname can be effectively applied to other LLMs, and better LLMs can enhance its performance.

In summary, we make the following contributions:

\begin{itemize}[leftmargin=*]
\item We provide insights into how large language models (LLMs) can enhance the performance of the SZZ algorithm while also highlighting the limitations in this task.
\item Based on these insights, we propose a novel approach to fully leverage the LLM's capabilities, which consists of two methods for locating bug-inducing commits, with the choice of method being adaptive to the LLM's ability to comprehend the bug.

\item We implement \appname on two popular programming languages and evaluate it on three developer-annotated datasets. The experimental results show that \appname outperforms all other baselines across the datasets.

\end{itemize}

%% file: tex/background.tex
In this section, we first introduce the SZZ algorithm's variants. Then we present our motivation examples.

\subsection{SZZ algorithms}\label{sec:szz}

\noindent\textbf{AG-SZZ.} The AG-SZZ algorithm was proposed by Kim et al.~\cite{kim2006automatic}. They observed that some changes in bug-fixing commits, such as blank lines, comments, and cosmetic changes, do not affect the program's behavior. Therefore, they excluded these changes when tracing back deleted lines. Additionally, they utilized the annotation graph instead of simply using the annotate command, as the annotation graph provides more detailed information about line changes and movements.

\noindent\textbf{MA-SZZ.} Da Costa et al. proposed the MA-SZZ algorithm.~\cite{da2016framework}. They found that the AG-SZZ algorithm mistakenly identifies commits with only meta-changes as bug-inducing commits. Meta-changes refer to branch changes, merge changes, and property changes. Da Costa et al. addressed this issue by connecting all meta-change nodes in the annotation graph to their prior changes, ensuring that the MA-SZZ algorithm does not include meta-changes as bug-inducing commits.

\noindent\textbf{R-SZZ and L-SZZ.} L-SZZ and R-SZZ, both based on the AG-SZZ algorithm, were proposed by Davies et al.~\cite{davies2014comparing}. They improved the AG-SZZ algorithm by selecting only one commit as the bug-inducing commit from the results produced by the AG-SZZ algorithm. R-SZZ selects the commit with the most recent date, while L-SZZ selects the commit with the most changed lines.

\noindent\textbf{RA-SZZ.} Neto et al.~\cite{neto2018impact} proposed the RA-SZZ algorithm after discovering that previous SZZ algorithms trace back changed lines related to refactoring operations when locating bug-inducing commits. Since refactoring operations do not affect the program's behavior, including them may introduce noise. Therefore, they used two tools RefDiff~\cite{silva2017refdiff} and Refactoring Miner~\cite{tsantalis2018accurate} to exclude refactoring modifications before tracing back lines. However, this algorithm is limited to Java projects, as the two tools mentioned above cannot work on other programming languages.%

\noindent\textbf{Neural-SZZ.} Neural-SZZ, proposed by Tang et al.~\cite{tang2023neural}, is based on deep learning. They observed that the previous methods fail to consider the semantic meaning of changed lines and the relationships between them. To address this, they utilize the CodeBERT~\cite{feng2020codebert} model to embed the changed lines, capturing their semantic meanings.  Additionally, they use a heterogeneous graph attention network (HAN)~\cite{wang2019heterogeneous} to capture the relationships between changed lines. After obtaining the embeddings of the changed lines, they employ the RankNet~\cite{burges2010ranknet} model to select the deleted lines that are most likely to be the root cause of the bug. Finally, they trace back the top N lines in the ranked list to locate bug-inducing commits. The authors implemented the algorithm only for the Java programming language.

\subsection{Potential and limitations of LLMs}~\label{sec:motivation}

\begin{figure}[t]
  \centering
  \includegraphics[width=0.8\linewidth]{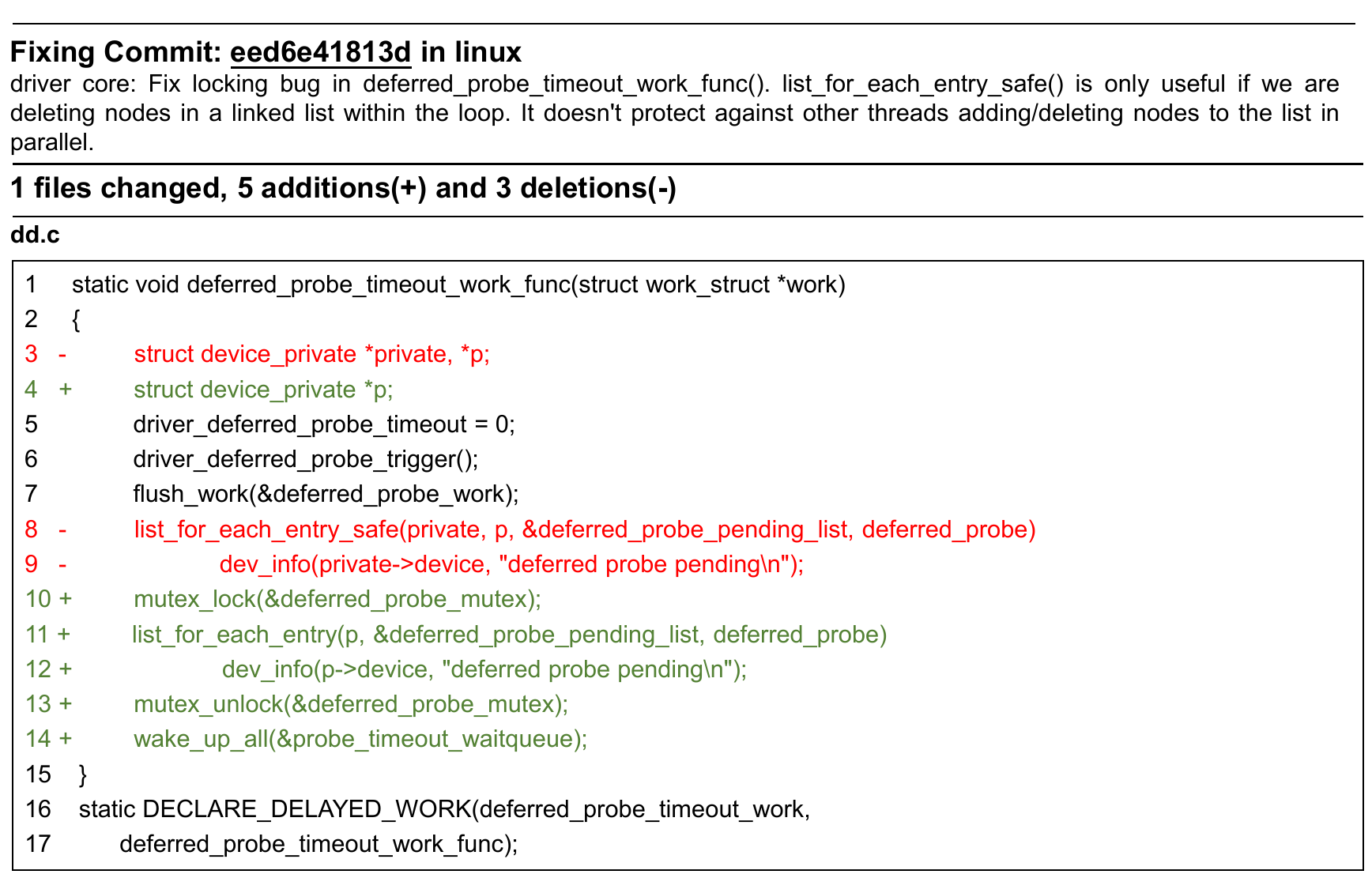}
  \vspace{-2mm}
  \caption{Motivation example one}\label{fig:motivation1}
  \vspace{-5mm}
\end{figure}

In this subsection, we present motivation examples to demonstrate the potential and limitations of LLMs in locating bug-inducing commits. We utilize the LLM GPT-4o-mini~\cite{gpt-4o-mini} to illustrate these examples.

\textbf{LLMs have the potential to identify the root cause of the bug from the bug-fixing commit and reduce false positives by pinpointing the bug-inducing commit from a set of candidates.} %
We illustrate this with the example presented in Figure~\ref{fig:motivation1}, which involves a bug-fixing commit $eed6e41813d$ in Linux.
We feed the prompt, "Based on the content of the bug-fixing commit, analyze the root cause of the bug and output the code statements leading to the bug", along with the content of the bug-fixing commit to the LLM. The LLM successfully predicts that the bug occurs because the function
\code{list\_for\_each\_entry\_safe} fails to protect the list when multiple threads add or delete nodes in parallel. It identifies lines 8 and 9 as buggy statements, filtering out line 3. Tracing back these two lines will yield two candidate bug-inducing commits, $eb7fbc9fb11$ and $25b4e70dcce$. We then use the LLM to determine which candidate commit introduces the bug. The LLM finds that the commit $eb7fbc9fb11$ introduces line 9 but only modifies the second parameter of the \code{dev\_info} function, which does not affect the existence of the bug.  Consequently, we filter out commit $eb7fbc9fb11$ and identify $25b4e70dcce$ as the final bug-inducing commit.

\begin{figure}[t]\vspace{-3mm}
  \centering
  \includegraphics[width=\linewidth]{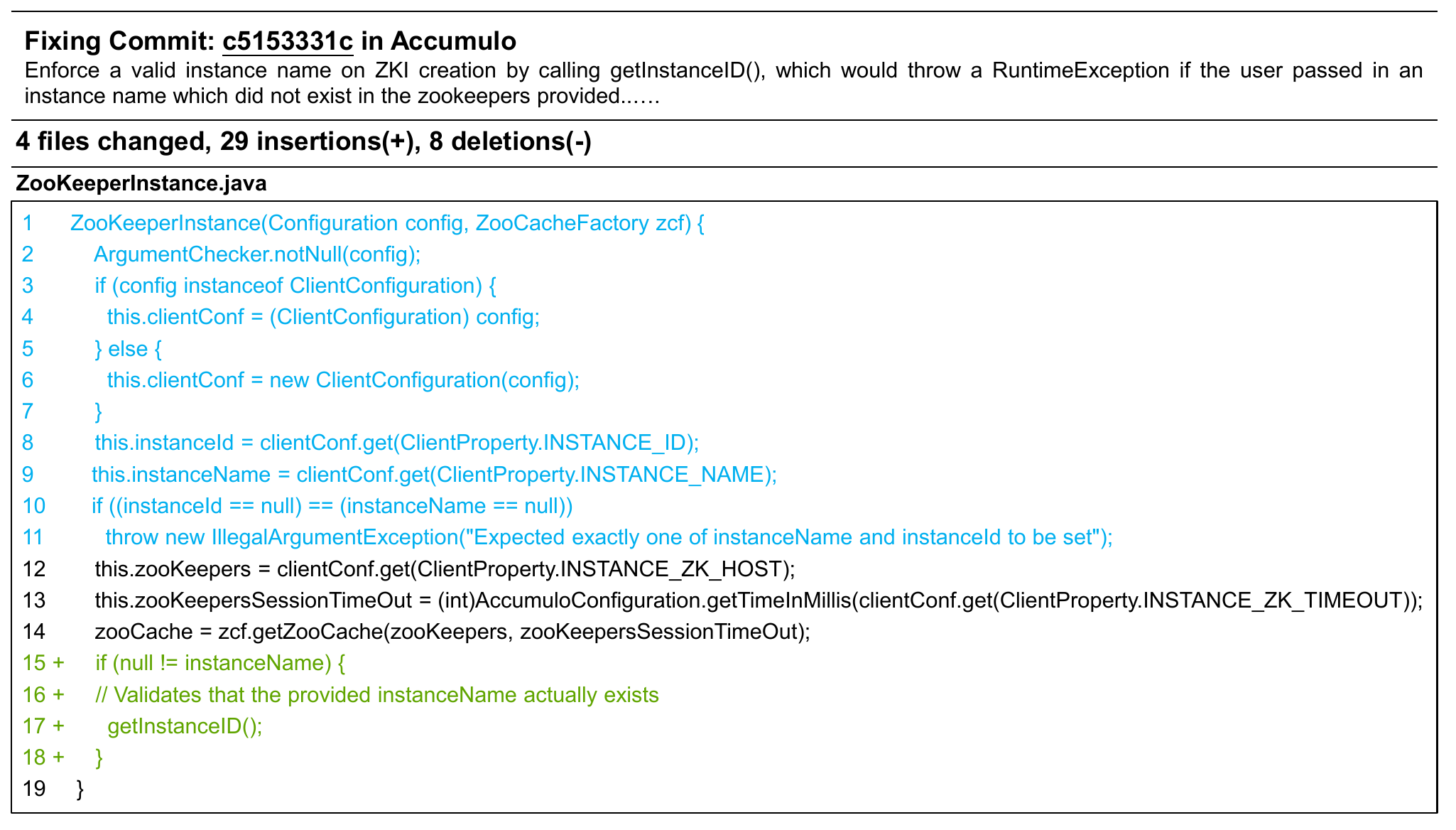}
  \vspace{-7mm}
  \caption{Motivation example two}\label{fig:motivation2}
  \vspace{-6mm}
\end{figure}

\textbf{However, LLMs face challenges when handling large bug-fixing commits, so it is beneficial to filter out irrelevant files before identifying buggy statements.}
This is demonstrated in the second motivation example illustrated in Figure~\ref{fig:motivation2}. 
This commit modifies four files, introducing twenty-nine insertions and making eight deletions. Due to page limits, we only show the most important part related to the identification of bug-inducing commits.
The lines highlighted in blue are added by us and are not in the original patch content. %
According to the commit message, the bug arises because the program fails to call the \code{getInstanceId} function to enforce a valid instance name and the bug is only related to the \code{instanceName} variable. If we directly input the whole patch into the LLM and require it to identify the code statements leading to the bug, it erroneously points the \code{@Test(expected = RuntimeException.class)} statement in another file named ZooKeeperInstanceTest.java.

\textbf{This example also demonstrates that LLMs have the potential to understand commit messages and accurately locate buggy statements, but they need sufficient context.} If we exclude the other files and only feed the LLM with changes in the correct file ZooKeeperInstance.java, the LLM still cannot output the correct code statements. Concretely, if we provide the LLM with the commit message and the original patch content (lines 12 to 19 in Figure~\ref{fig:motivation2}), it still incorrectly identifies line 14 as buggy code statements. This is due to insufficient context.
According to the commit message, the bug is related to the variable \code{instanceName}. However, in the original patch content(lines 12 to 19), the only code statement related to the variable \code{instanceName} is line 15, which is used to fix the bug. The full content of the \code{ZooKeeperInstance constructor} (lines 1 to 19 in Figure~\ref{fig:motivation2})  contains the statement \code{this.instanceName = clientConf.get(ClientProperty.INSTANCE\_NAME)}, which relates to the \code{instanceName} variable. But this statement is not displayed in the original patch. By providing the expanded context(lines 1 to 19), which includes the entire constructor, the LLM can correctly identify the code statement in line 9.

%% file: tex/Approach.tex
\begin{figure}[t] \vspace{-3mm}
  \centering
  \includegraphics[width=0.8\linewidth]{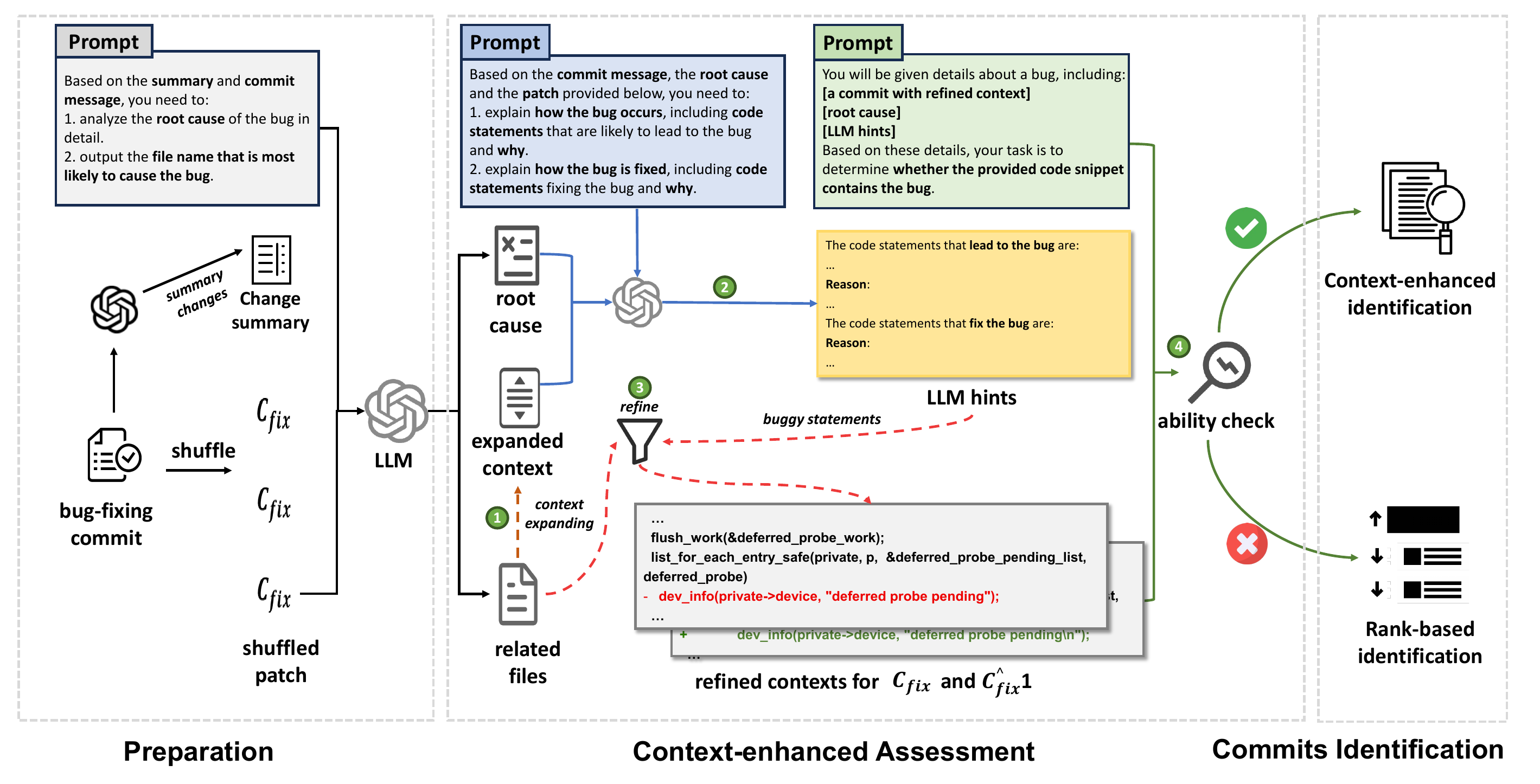}
  \vspace{-5mm}
  \caption{Overview of \appname}\label{fig:overview}
  \vspace{-7mm}
\end{figure}

Building on the motivation examples, we propose a new framework called \appname to effectively detect buggy statements and locate bug-inducing commits. Fig.~\ref{fig:overview} presents the overview of our framework, which consists of three parts: preparation, context-enhanced assessment, and commits identification. 
In the preparation phase, we analyze the bug-fixing commit, identify the core files related to the bug, and determine its root cause. During the context-enhanced assessment, we assess whether the LLM can understand the bug and determine its presence in the commit. If the LLM demonstrates this ability, we employ the context-enhanced identification approach during the commit identification process; otherwise, we fall back to the rank-based identification approach.

\subsection{Preparation}\label{Preparation}

In this step, we use the large language model (LLM) to analyze bug-fixing commits. We aim to summarize the root cause of the bug based on the bug-fixing commit and filter out irrelevant files. 
In motivation example one, we have shown that irrelevant files undermine the LLM's ability and we need to filter them out. 

Following the chain-of-thought (CoT) concept~\cite{wei2022chain}, we first require the LLM to analyze the patch, summarizing the modifications and their interrelationships within the bug-fixing commit. 
Next, we ask the LLM to identify the root cause of the bug and the related files based on the modification summary and the commit message.

To enhance performance when handling large bug-fixing commits with multiple modified files, we employ two additional approaches. First, we shuffle the sequence of modified files in the patch, ensuring that each file has an equal chance of being identified as related to the root cause. Previous studies~\cite{liu2024lost} reveal that LLMs tend to ignore content in the middle of text when handling long texts. 
Second, we run the LLM three times for the same question and shuffle the patch at each run-time. This approach is similar to a voting system~\cite{wang2022self}. 
However, instead of only considering files with majority votes, we take a more conservative approach: if a file name appears in any of the LLM outputs, we regard it as related to the root cause. This strategy helps minimize the risk of omitting important files. After this step, we obtain the root cause of the bug and filter out all irrelevant files.

\subsection{Context-enhanced assessment} \label{sec:Context-enhanced-assessment}

In this section, we explain the necessity of the assessment and our approach to it. 
In the first motivation example, we demonstrate that LLMs can determine whether a bug exists in a commit, allowing us to use them to select the final bug-inducing commit from a set of candidates. The second example illustrates that providing more context can enhance the LLM's ability to understand the patch and help identify buggy statements more accurately. However, previous studies in automatic program repair have shown that LLMs still do not comprehend certain bugs~\cite{parasaram2024fact,bouzenia2024repairagent}, even with enough context. This indicates that LLMs are unable to determine whether these kinds of bugs exist in programs because they cannot understand the bugs. If LLMs cannot understand the bug even with additional context, providing more context becomes meaningless, and alternative methods are required to address these cases. Therefore, it is crucial to assess the LLM's ability to comprehend the bug and identify its presence.

To assess the LLM’s ability to determine whether the bug exists, we need two versions of the program: one containing the bug and another where it has been fixed.  Therefore, an ideal approach is to make use of the bug-fixing commit ${C_{fix}}$, where the version ${C_{fix}}^{\wedge}1$ is buggy and version ${C_{fix}}$ is correct. 
Although we can directly require the LLM to assess whether the commit is buggy based on the root cause of the bug, experimental results indicate that this approach yields low performance (see section~\ref{tab:rq3-result}).  Instead, we first require the LLM to generate a hint to assist in determining whether the commit contains the bug. The hint includes detailed information about the code statements in the patch. Its further specifics will be provided later in this section.
Then, we separately feed the hint and the bug-related contexts extracted from commit ${C_{fix}}$ and ${C_{fix}}^{\wedge}1$ to the LLM, asking it to identify whether each version contains the bug. If the LLM even cannot identify the two versions correctly using its own produced hint, we regard that the LLM is unable to comprehend the bug, let alone select the final bug-inducing commit from a set of candidates.

As shown in Figure~\ref{fig:overview}, the entire context-enhanced assessment process consists of four steps, which are as follows:

\ding{182} \textbf{Context Expanding}: First, we provide the LLM with sufficient context through a process that we call context expanding. Previous studies have found that the contextual code is crucial for providing information to the model~\cite{chen2019sequencer, xia2023revisiting}. However, the bug-fixing commit often does not contain the full content of the changed functions. 
The partial content of functions in the fixing commit may hinder the LLM's ability to understand both the functions and the modifications.
Therefore, for each modified function, we expand its context by extracting the full content of both buggy and fixed versions and generating their diffs.
For modified lines outside the function, we expand their context by extracting three unmodified lines around them. 
We have presented an example of context expanding in the second motivation example, as illustrated in Figure~\ref{fig:motivation2}.
Specifically, the code highlighted in blue represents new additions, and the others are collected from the original patch. As indicated in Section~\ref{sec:motivation}, the buggy code \code{this.instanceName = clientConf.get(ClientProperty.INSTANCE\_NAME)} is not located in the original patch but in the expanded context, indicating the necessity of the context expanding.

 {\ding{183} \textbf{Hint Generation}}: 
Next, we require the LLM to establish a hint to determine whether the bug exists in a commit. Specifically, we ask the LLM to identify the code statements leading to the bug and provide a reason. Additionally, we require the LLM to identify the code statements that fix the bug and provide a reason. Note that we do not limit the LLM to choosing code statements only from deleted lines. It can select any code statements from the expanded context.

{\ding{184} \textbf{Context Refinement}: Before assessing the LLMs' ability to determine whether the bug exists in a commit, we need to refine the expanded context to obtain the refined context. This step is necessary because the expanded context may contain much irrelevant content that is not related to the bug. For example, the expanded context might include an entire function with hundreds of lines, while only a few lines are relevant to the bug. Feeding the expanded context directly to LLMs may undermine their ability to assess the existence of the bug in a commit, as previous studies~\cite{li2023loogle} suggest that LLMs struggle with intricate tasks when handling long texts. Therefore, we attempt to refine the expanded context. We first extract the buggy statements identified in the hint from the file in commit ${C_{fix}}^{\wedge}1$. These buggy statements are then sorted in ascending order based on their line numbers \{$l_1$, $l_2$, ... $l_n$ \} in commit ${C_{fix}}^{\wedge}1$, where $l_1$ is the smallest line number and $l_n$ is the largest. Here, we define $l_{min}$ as $l_1-N$ and $l_{max}$ as  $l_n+N$. N is a constant starting from 3 to ensure that lines $l_{min}$ and $l_{max}$ can be mapped to corresponding lines in commit $C_{fix}$. If line $l_{min}$ or line $l_{max}$ cannot be mapped, we keep incrementing N. Then, we extract the content ranging from the line number $l_{min}$ to the line number $l_{max}$, forming the refined context for commit ${C_{fix}}^{\wedge}1$. To obtain the refined context for commit ${C_{fix}}$, we map lines $l_{min}$  and  $l_{max}$ to their corresponding line numbers $l_{min}^{'}$ and $l_{max}^{'}$ and extract the content between these two line numbers in commit ${C_{fix}}$, forming the refined context for commit ${C_{fix}}$.

{\ding{185} \textbf{Ability Check}: Finally, with the contexts for both commits obtained, we begin to check the LLM's ability to determine whether the bug exists in the commit. We provide the LLM with the root cause of the bug, the hint collected above, and the refined contexts for two versions. If the LLM can correctly identify the fixed version and the buggy version, we proceed to adopt context-enhanced identification. Otherwise, we fall back to rank-based identification.

\begin{figure}[t]
  \centering
  \includegraphics[width=0.80\linewidth]{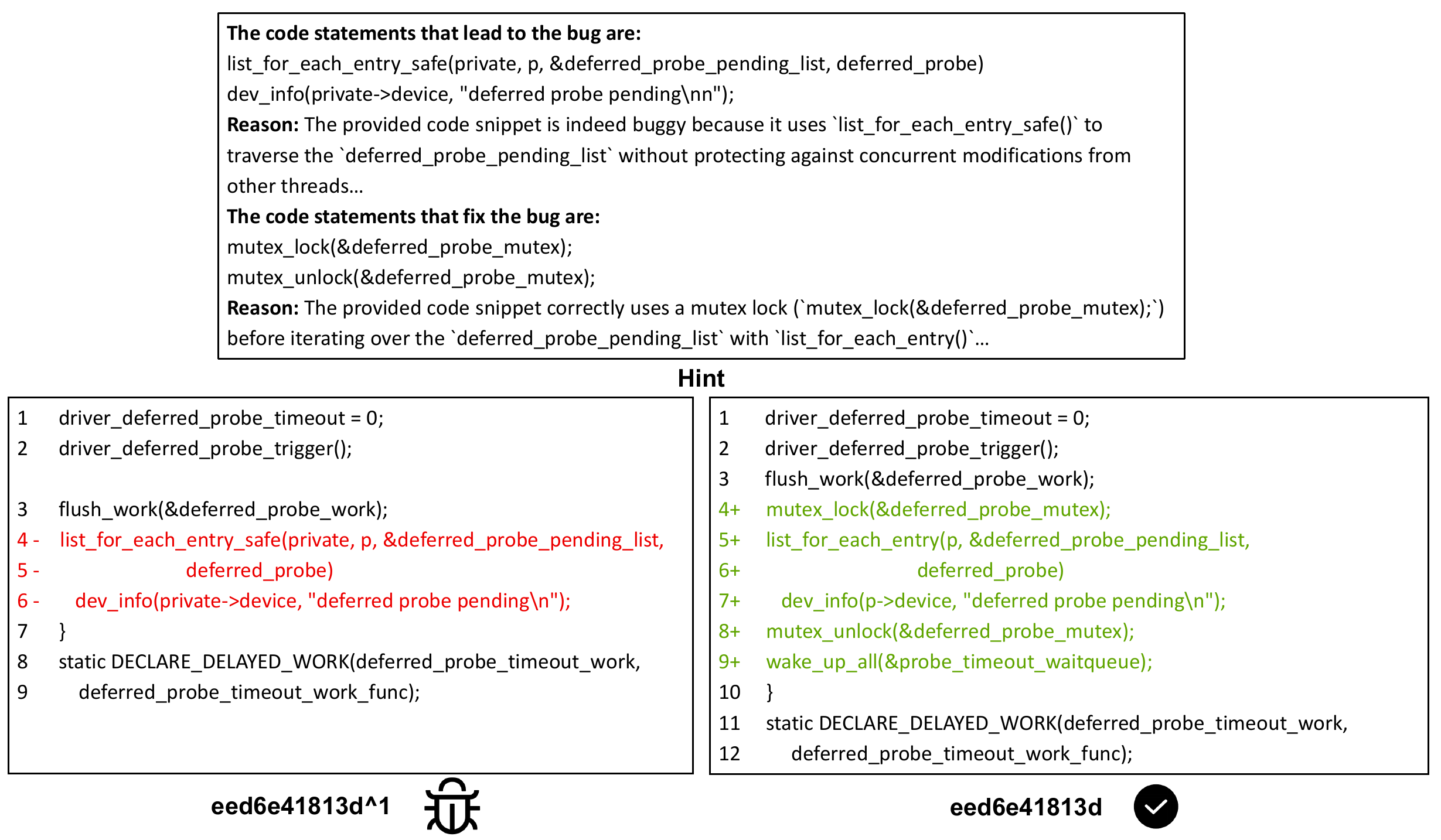}
  \caption{An example of context-enhanced ability check}\label{fig:strategy-selection}
  \vspace{-5mm}
\end{figure}

One example of the context-enhanced ability check is shown in Figure~\ref{fig:strategy-selection}, which corresponds to the motivation example one. Here, we set N to 3. The LLM identifies lines from 4 to 6 as buggy statements in ${eed6e41813d}^{\wedge}1$. Therefore, $l_{min}$ is 1 and $l_{max}$ is 9. To generate the refined context for this commit, we extract lines ranging from line 1 to line 9.  We then map line 1 and line 9 to commit ${eed6e41813d}$, getting $l_{min}^{'}$ as 1 and $l_{max}^{'}$ as 12.  Finally, we extract the lines between them, forming the refined context in commit ${eed6e41813d}$. We feed the LLM with the root cause of the bug, the hint, and two versions of the context. It identifies the context for commit ${eed6e41813d}^{\wedge}1$ as buggy and the context for commit ${eed6e41813d}$ as correct, demonstrating its ability to determine whether the bug exists in a commit.

\subsection{Commits identification}

\subsubsection{Context-enhanced identification}

\begin{figure}[t]
  \centering
  \includegraphics[width=0.60\linewidth]{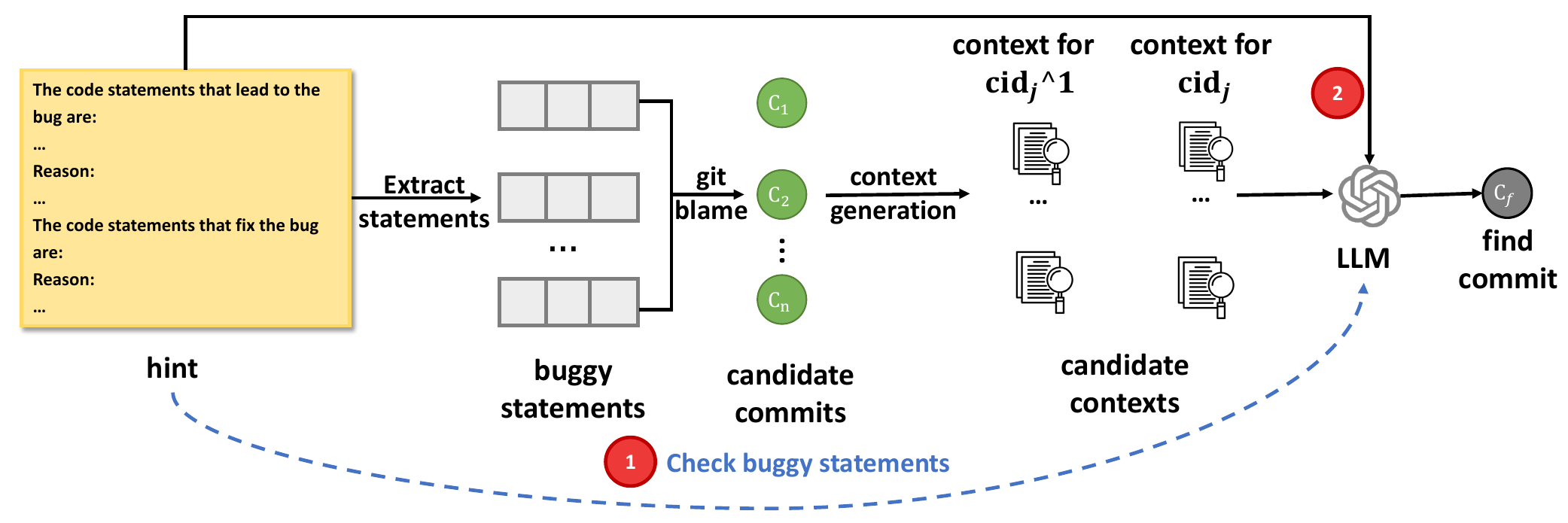}
  \caption{The workflow of context-enhanced identification}\label{fig:strategy2-workflow}
  \vspace{-3mm}
\end{figure}

Given a bug-fixing commit, once we verify that the LLM can understand the bug in it with the expanded and refined context, we apply \textbf{context-enhanced identification} to identify the bug-inducing commits, as shown in Figure~\ref{fig:strategy2-workflow}.
First, we retrieve all buggy statements from the hint. We then trace back these buggy code statements, obtain a set of candidate commits, and sort them in descending order by commit date, forming a candidate list \{$C_1$, $C_2$, ... $C_i$\}. Next, we generate the refined context for each candidate commit following the same process used in the ability check. For each candidate commit $C_j$, we create the context for both $C_j$ and its previous version $C_j^{\wedge}1$.

To enhance the LLM’s ability to determine whether the candidate commit $C_j$ contains the bug,
we split the determination process into two steps. In the first step, we input the context of commit
$C_j$ and the buggy statements in the hint to the LLM, asking the LLM to determine whether the
commit contains the buggy code statements or code statements with similar semantic meanings to
the buggy code statements. If the LLM answers "no", we simply believe that the commit $C_j$ does
not contain the bug. If the LLM answers "yes", we proceed to let it determine whether the commit
is buggy. In the second step, we feed the context of commit $C_j$
, the root cause of the bug, and the
hint to the LLM, asking it to determine whether the commit contains the bug.

We utilize the LLM to check the candidate commits in the list from index $1$ to $i$. If we can find an index $f$, where the LLM believes that the context of the commit $C_f$ is buggy but $C_f^{\wedge}1$ is not, we designate it as the bug-inducing commit. If we can not find such a commit, we fall back to rank-based identification conservatively.

\begin{figure}[t]
  \centering
  \includegraphics[width=0.80\linewidth]{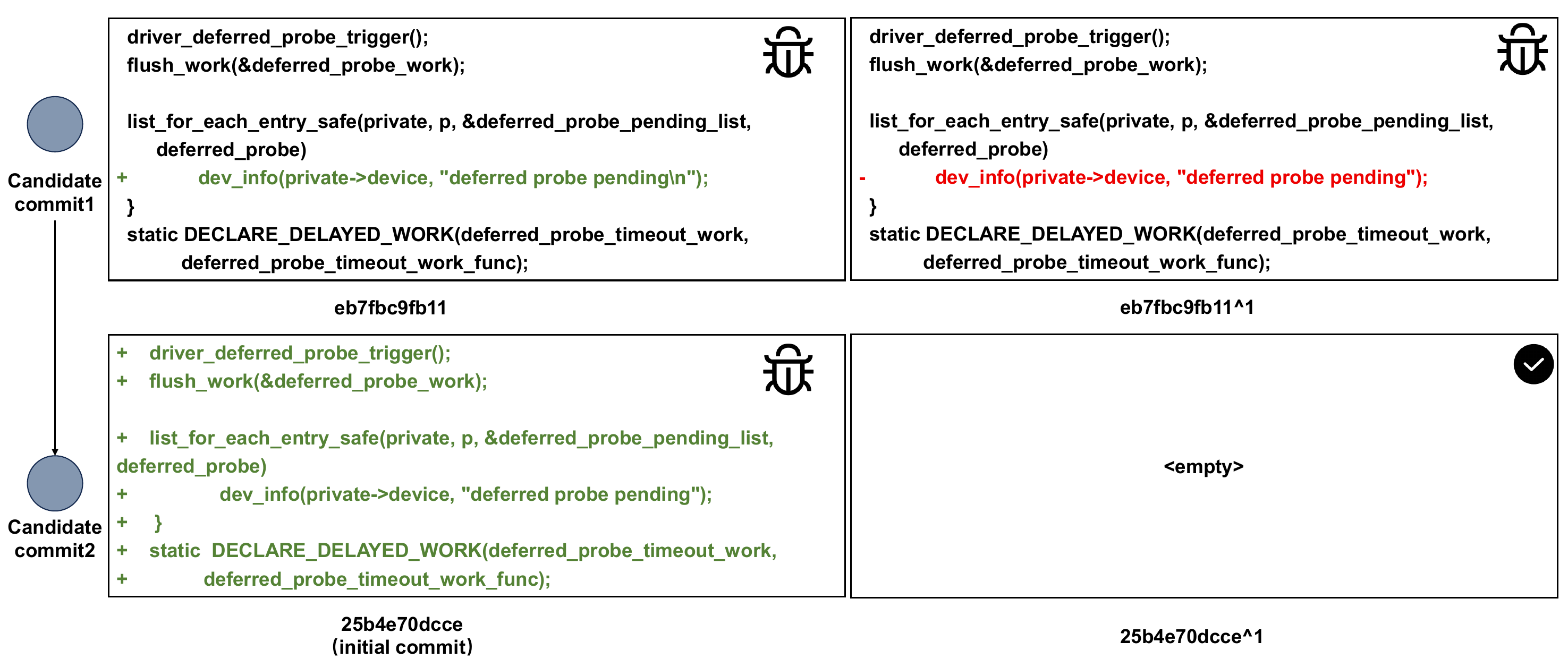}
  \caption{An example of context-enhanced identification}\label{fig:strategy2-example}
  \vspace{-5mm}
\end{figure}

Figure~\ref{fig:strategy2-example} provides an example, 
which corresponds to the first motivation example. Firstly, \appname traces back the buggy statements in the hint and finds two candidate commits, the commit $eb7fbc9fb11$ (denoted as $C_1$) and commit $25b4e70dcce$ (denoted as $C_2$). We sort them based on their commit date in descending order and get the candidates list \{$C_1$, $C_2$\}. We first check the contexts for commits $C_1$ and ${C_1}^{\wedge}1$, following the steps above and requiring the LLM to determine whether the two versions are buggy. The LLM identifies that both the $C_1$ and ${C_1}^{\wedge}1$ versions of the program contain the bug. Therefore, this candidate commit is not the bug-inducing commit. \appname then checks the commit  $C_2$ and ${C_2}^{\wedge}1$ , following the same steps. Here, $C_2$ is the initial commit that introduces the file. Therefore, the context of ${C_2}^{\wedge}1$ is empty. The LLM finds the commit $C_2$ contains the bug and the context of commit ${C_2}^{\wedge}1$  contains no code statements and is not buggy. Therefore, \appname finally designates the commit $C_2$ as the final bug-inducing commit.

\subsubsection{Rank-based identification}

\textbf{Rank-based identification} addresses cases where the LLM cannot fully understand the bug and cannot determine whether the bug exists in the program. Therefore, in this approach, we do not use LLMs to select the final bug-inducing commit from candidate commits. Instead, we simply follow the idea of NerualSZZ~\cite{tang2023neural} and proceed as follows:

\begin{itemize}
\item \textbf{Buggy Statements Identification:} We first ask the LLM to identify buggy statements from the bug-fixing commit based on the commit message and the root cause obtained in section \ref{Preparation}. In this phase, we input only the root cause, the commit message, and the original changed files obtained in section~\ref{Preparation}. We do not provide the LLM with expanded context, as we observe that if the LLM cannot understand the bug, additional context will undermine its performance (see Section~\ref{rq2}).
    \item \textbf{Relevance Ranking:} By utilizing a listwise rank algorithm~\cite{sun2023chatgpt} based on the LLM, we rank these buggy statements according to their relevance to the root cause.
    \item \textbf{Candidate Commits generation:} For each file, we retrieve the top N code statements, trace them back to their corresponding commits, and add these commits to our list of candidate commits.
    \item \textbf{Final Commit Designation:} We then sort these candidate commits by their commit date and designate the most recent commit as the bug-inducing commit. This approach aligns with previous studies~\cite{rodriguez2020bugs,bao2022v}, which suggest that bugs are typically introduced by recent commits.
\end{itemize}

%% file: tex/experiment_set_up.tex
\subsection{Dataset}

\begin{table*}[!t]
\caption{The statistics of the bugs and corresponding bug fixing commits in three datasets}\label{tab:dataset}
\resizebox{0.55\textwidth}{!}{
\begin{tabular}[c]{llrrrr}
\toprule
\textbf{Dataset}  & \textbf{Project} & \textbf{\#Bug-Fixing} &  \textbf{\#Bug-Inducing} & \textbf{\#SMALL} & \textbf{\#LARGE} \\

\midrule
{\sc \textbf{DS\_LINUX}} & \makecell[l]{ linux} & \makecell[r]{ 1,500  } & \makecell[r]{ 1,562  }  &\makecell[r]{681}  &\makecell[r]{819}   \\

\midrule
{\sc \textbf{DS\_GITHUB}} & \makecell[l]{ systemd\\ qemu \\ gpac \\ unitime \\ JohnTheRipper \\ libvirt \\ opensips \\ ...(279 more projects)} & \makecell[r]{ 15 \\ 10 \\ 9 \\ 6 \\ 5 \\4 \\ 4 \\308} & \makecell[r]{ 15 \\ 10 \\ 9 \\ 6 \\ 5 \\4 \\4 \\309}  & \makecell[r]{ 5 \\ 3 \\ 2 \\ 2 \\ 2 \\2 \\1 \\ 130} & \makecell[r]{ 10 \\ 7 \\ 7 \\ 4 \\ 3 \\2 \\3 \\ 171} \\
\cmidrule{2-6}
  & \makecell[l]{Total} & 361 & 362 & 146 & 207  \\

\midrule
{\sc \textbf{DS\_APACHE}} & \makecell[l]{ accumulo\\ ambari \\ hadoop \\ lucene \\ oozie} & \makecell[r]{ 35 \\ 38 \\ 53 \\ 70 \\ 45} & \makecell[r]{ 55 \\ 44 \\ 57 \\ 145 \\ 50} & \makecell[r]{ 7 \\ 1 \\ 6 \\ 3 \\ 3}  & \makecell[r]{ 28 \\ 37 \\ 47 \\ 68 \\ 42}  \\ 
\cmidrule{2-6}
  & \makecell[l]{Total} & 241 & 351 &20 & 222  \\

\bottomrule
\end{tabular}

}
\end{table*}

To evaluate our method, we require high-quality datasets containing bug-fixing commits and their corresponding bug-inducing commits. Previous research~\cite{wen2019exploring} has demonstrated that datasets produced by the SZZ algorithm contain significant noise, leading us to discard these datasets. Other available datasets are annotated by researchers~\cite{neto2019revisiting, davies2014comparing}. While these datasets are of higher quality, researchers may not have the same level of knowledge as developers about specific projects, which still can result in inaccuracies. To ensure accuracy, we combined three developer-annotated datasets to form the final dataset for evaluating our method. In these datasets, all bug-fixing commits and bug-inducing commits are annotated by developers, and are extracted from bug reports or commit messages.

\textbf{DS\_LINUX} refers to the dataset created by Lyu et al. ~\cite{lyu2024evaluating}, which is based on the Linux kernel.
The researchers observed that Linux developers label bug-fixing commits with their corresponding bug-inducing commits in the commit messages. They collected these commit messages and built the dataset based on them. This dataset is notable for its size, containing 76,046 pairs of bug-fixing and bug-inducing commits. However, its drawback is that it is only related to the Linux kernel.

\textbf{DS\_GITHUB} refers to the dataset constructed from multiple repositories on GitHub, collected by Rosa et al.~\cite{rosa2021evaluating}.
The authors mined GitHub by first locating the bug-fixing commits and then identifying the corresponding bug-inducing commits based on the information left by developers in the commit messages. 
This dataset is characterized by its inclusion of hundreds of repositories. However, its drawback is that each repository contains very few bug-fixing commits. Moreover, this dataset includes many repositories with few stars. 

\textbf{DS\_APACHE} refers to the dataset created from several Apache projects, collected by Wen et al.~\cite{wen2019exploring}.
The researchers extracted the bug reports from several Apache projects, obtaining bug-fixing commits and their corresponding bug-inducing commits based on the bug reports and commit messages. This dataset contains several Apache projects with high star ratings, and each project corresponds to a moderate number of bug-fixing commits.

Table~\ref{tab:dataset} presents the statistics of the three datasets. Since the majority of the combined datasets are comprised of C and Java projects, we include only C and Java projects in the final dataset. \begin{coloredtext}To control experimental costs, we sample data from DS\_LINUX following previous studies~\cite{xue2024automated,yang2024exploring,li2024only}, using a 95\% confidence level and a margin of error below 5\%.  We sample 1,500 bug-fixing commits from a total of 76,046 commits, along with their corresponding bug-inducing commits. This sample size is comparable to previous studies. For example, Li et al.~\cite{li2024only} sampled 381 commits from a total of 35,431 commits to evaluate their approach for generating commit messages. Note that we do not sample data from DS\_GITHUB and DS\_APACHE. 
We also provide information about the size of bug-fixing commits. Following previous studies~\cite{bao2022v,tang2023neural}, if a bug-fixing commit contains more than five changed lines, we categorize it as a large commit, otherwise, we categorize it as a small commit. From the table, we observe that the number of small bug-fixing commits is roughly equal to the number of large bug-fixing commits in DS\_LINUX and DS\_GITHUB, while most bug-fixing commits in DS\_APACHE are large.
In summary, our dataset comprises multiple high-quality pairs of bug-fixing and bug-inducing commits in various programming languages across numerous repositories.\end{coloredtext}

\subsection{Experiment Setting}\label{experiment-setting}

Our experiment is conducted on a server equipped with two NVIDIA A800 GPUs and an Intel Xeon 6326 CPU, running on Ubuntu OS. We utilize gitpython~\cite{GitPython} to extract patch content and obtain the necessary information about commits. Additionally, we use the \emph{tree-sitter}~\cite{treesitter} parser to extract functions from the source code when generating the context. Although our implementation focuses on the Java and C programming languages, \appname is generic and can be easily extended to other programming languages by altering the parser in \emph{tree-sitter}.

For LLMs, we aim to balance cost and effectiveness for proprietary models. Therefore, we choose GPT-4o-mini due to its low fees and relatively high effectiveness. We estimated that using GPT-4 would cost approximately \$300 per round, which is prohibitively expensive. Our experiments show that GPT-4o-mini is sufficient for our needs. For open-source LLMs, we use Llama3-8b and Llama3-70b, which we downloaded from Hugging Face~\cite{HuggingFace}.\begin{coloredtext} To reduce randomness, we set the temperature to 0.0 for both GPT-4o-mini and the open-source LLMs, aligning with settings used in previous studies~\cite{li2024only,xu2024unilog}. We also employ two strategies to further address randomness. First, we repeat the entire experiment three times and calculate the average metrics across the three runs. \end{coloredtext}   Second, our dataset consists of 2,102 test cases, which is large enough to reveal statistical patterns and minimize the influence of individual test cases on the final results.

We use the SZZ algorithms introduced in Section~\ref{sec:szz} as our baselines. Implementations of the B-SZZ, AG-SZZ, MA-SZZ, L-SZZ, R-SZZ, and RA-SZZ algorithms are from the replication package provided by Rosa et al.~\cite{rosa2021evaluating}. The implementation of the NeuralSZZ algorithm is from the replication package provided by Tang et al.~\cite{tang2023neural}. \begin{coloredtext}We train the model using the same training set as in the original paper and achieve nearly identical performance on its own test set. We then apply the trained model to DS\_GITHUB and DS\_APACHE-j. 
 \end{coloredtext} To evaluate our approach and baselines, we employ three widely used metrics: Precision, Recall, and F1-score, following the methodology used in previous studies~\cite{bao2022v, lyu2024evaluating}.

%% file: tex/experiments.tex
In this section, we first demonstrate the effectiveness of \appname (RQ1). Next, we evaluate the impact of its key components (RQ2). Finally, we show that \appname can be applied to other large language models (RQ3).

\subsection{RQ1. Effectiveness of \appnamebold in identifying bug-inducing commits}\label{RQ1}
\begin{table*}[!t]\vspace{-5mm}
\caption{The performance comparisions between all methods in finding c bug-inducing commits}\label{tab:rq1-result1}
\vspace{-3mm}
\resizebox{0.50\textwidth}{!}{

\begin{tabular}[c]{lccccccccc}
\toprule
\multirow{2}{*}{\textbf{Method}} & \multicolumn{3}{c}{\textbf{DS\_LINUX}}  & \multicolumn{3}{c}{\textbf{DS\_GITHUB-c}}   \\

& \makecell[l]{Precision} & \makecell[r]{Recall} & \makecell[r]{F1-score} & \makecell[l]{Precision} & \makecell[r]{Recall} & \makecell[r]{F1-score}  \\
\midrule

\makecell[l]{B-SZZ} & \makecell[c]{0.452} & \makecell[c]{\textbf{0.578}} & \makecell[c]{\underline{0.507}} & \makecell[c]{0.361} & \makecell[c]{\textbf{0.656}} & \makecell[c]{0.466}  \\
\midrule

\makecell[l]{AG-SZZ} & \makecell[c]{0.448} & \makecell[c]{0.553} & \makecell[c]{0.495} & \makecell[c]{0.410} & \makecell[c]{0.592} & \makecell[c]{0.484} \\
\midrule

\makecell[l]{MA-SZZ} & \makecell[c]{0.421} & \makecell[c]{0.538} & \makecell[c]{0.472} & \makecell[c]{0.335} & \makecell[c]{0.624} & \makecell[c]{0.436}  \\
\midrule

\makecell[l]{R-SZZ} & \makecell[c]{0.583} & \makecell[c]{0.448} & \makecell[c]{\underline{0.507}} & \makecell[c]{0.671} & \makecell[c]{0.582} & \makecell[c]{\underline{0.620}}  \\
\midrule

\makecell[l]{L-SZZ} & \makecell[c]{0.560} & \makecell[c]{0.430} & \makecell[c]{0.486} & \makecell[c]{0.486} & \makecell[c]{0.422} & \makecell[c]{0.452} \\
\midrule

\makecell[l]{\appname} & \makecell[c]{\textbf{0.628}} & \makecell[c]{0.552} & \makecell[c]{\textbf{0.588}} & \makecell[c]{\textbf{0.687}} & \makecell[c]{{0.641}} & \makecell[c]{\textbf{0.663}}  \\

\bottomrule
\end{tabular}

}
\vspace{-3mm}
\end{table*}

\begin{table*}[!t]
\caption{The performance comparison between methods in finding java bug-inducing commits}\label{tab:rq1-result2}
\vspace{-3mm}
\resizebox{0.50\textwidth}{!}{

\begin{tabular}[c]{lccccccccc}
\toprule
\multirow{2}{*}{\textbf{Method}} & \multicolumn{3}{c}{\textbf{DS\_GITHUB-j}}  & \multicolumn{3}{c}{\textbf{DS\_APACHE}}   \\ 

& \makecell[l]{Precision} & \makecell[r]{Recall} & \makecell[r]{F1-score} & \makecell[l]{Precision} & \makecell[r]{Recall} & \makecell[r]{F1-score}  \\
\midrule

\makecell[l]{B-SZZ} & \makecell[c]{0.285} & \makecell[c]{\textbf{0.680}} & \makecell[c]{0.401} & \makecell[c]{0.251} & \makecell[c]{\textbf{0.435}} & \makecell[c]{0.318}  \\
\midrule

\makecell[l]{AG-SZZ} & \makecell[c]{0.421} & \makecell[c]{0.533} & \makecell[c]{0.470} & \makecell[c]{0.328} & \makecell[c]{0.310} & \makecell[c]{0.318}  \\
\midrule

\makecell[l]{MA-SZZ} & \makecell[c]{0.239} & \makecell[c]{0.560} & \makecell[c]{0.335} & \makecell[c]{0.307} & \makecell[c]{0.345} & \makecell[c]{0.329}  \\
\midrule

\makecell[l]{R-SZZ} & \makecell[c]{0.538} & \makecell[c]{0.467} & \makecell[c]{0.500} & \makecell[c]{0.497} & \makecell[c]{0.288} & \makecell[c]{0.364}  \\
\midrule

\makecell[l]{L-SZZ} & \makecell[c]{0.492} & \makecell[c]{0.427} & \makecell[c]{0.457} & \makecell[c]{0.366} & \makecell[c]{0.211} & \makecell[c]{0.267}  \\
\midrule

\makecell[l]{RA-SZZ} & \makecell[c]{0.337} & \makecell[c]{0.440} & \makecell[c]{0.382} & \makecell[c]{0.264} & \makecell[c]{0.325} & \makecell[c]{0.293}  \\
\midrule

\makecell[l]{Neural-SZZ} & \makecell[c]{0.556} & \makecell[c]{0.486} & \makecell[c]{\underline{0.520}} & \makecell[c]{0.563} & \makecell[c]{0.364} & \makecell[c]{\underline{0.442}}  \\

\midrule

\makecell[l]{\appname} & \makecell[c]{\textbf{0.607}} & \makecell[c]{0.569} & \makecell[c]{\textbf{0.587}} & \makecell[c]{\textbf{0.610}} & \makecell[c]{0.398} & \makecell[c]{\textbf{0.482}}  \\

\bottomrule
\end{tabular}

\vspace{-8mm}
}
\end{table*}

Table~\ref{tab:rq1-result1} and~\ref{tab:rq1-result2} present the results of \appname and baselines in identifying bug-inducing commits in C and Java projects, respectively. 
Since DS\_GITHUB consists of both c projects and Java projects, we split it into  DS\_GITHUB-c and  DS\_GITHUB-j, containing C projects and Java projects, respectively.

As shown in Table~\ref{tab:rq1-result1}, for DS\_LINUX, all baselines perform almost the same, mirroring the experimental results in the original whole dataset~\cite{lyu2024evaluating}. This suggests that our selected dataset has the same statistical patterns as the original dataset. 
The B-SZZ and R-SZZ algorithms achieve the highest F1-scores among all baselines. Specifically, B-SZZ achieves the highest recall, while R-SZZ achieves the highest precision. 
\begin{coloredtext}
We also analyze why B-SZZ outperforms its variants, such as AG-SZZ, MA-SZZ, and L-SZZ in DS\_LINUX. B-SZZ outperforms AG-SZZ because Linux contains numerous bug-fixing commits related only to comments and configurations. While B-SZZ successfully identifies these commits, AG-SZZ filters them out. Similarly, MA-SZZ assumes that commits with only meta-changes are not bug-inducing, but DS\_LINUX shows that developers do label such commits as bug-inducing. L-SZZ, on the other hand, assumes the commit with the most changed lines among those identified by AG-SZZ is the bug-inducing commit. However, this assumption also often fails, as shown by the dataset. These limitations explain why these variants perform worse than the B-SZZ algorithm. \end{coloredtext}
In DS\_GITHUB-c, the R-SZZ algorithm performs the best, significantly outperforming all other baselines, with an F1-score of 0.620. \appname achieves the highest precision and F1-score across these two datasets. In DS\_LINUX, it improves precision by 7.7\% and F1-score by 16.0\% compared to the best baseline. In DS\_GITHUB-c, it improves precision by 2.4\% and F1-score by 6.9\%, respectively. 

From Table~\ref{tab:rq1-result2}, we observe that the NeuralSZZ algorithm performs the best in precision and F1-score among all baselines. This suggests that utilizing deep learning to rank code statements is an effective way to enhance performance. \appname also performs the best in precision and F1-score in these two datasets. Specifically, it improves precision by 9.2\% in DS\_GITHUB-j and 8.3\% in DS\_APACHE. Additionally, it enhances the F1-score by 12.9\% in DS\_GITHUB-j and 9.0\% in DS\_APACHE. \begin{coloredtext}
This demonstrates the effectiveness of \appname in handling large bug-fixing commits, as most bug-fixing commits in DS\_APACHE are large.\end{coloredtext}

Combining the three datasets, we observe that the R-SZZ algorithm performs better than other baselines except the Neural-SZZ algorithm. This is consistent with the finding of Rodriguez et al.~\cite{rodriguez2020bugs} that defects are typically introduced in the most recent changes. Moreover, we can find that all baselines' performance varies a lot in different datasets. For example, R-SZZ outperforms other baselines a lot in DS\_GITHUB-c but it performs almost the same as B-SZZ in DS\_LINUX. Neural-SZZ, the deep-learning based approach, also has the same problem. It outperforms other baselines a lot in DS\_APACHE but only shows a slight improvement over R-SZZ in DS\_GITHUB-j. 
In contrast, \appname does not have the same problem. In the worst scenario, it can still outperform other baselines by 6.9\% in  F1-score. Additionally, our method improves precision and F1-score without sacrificing too much recall. In all three datasets, \appname achieves a higher recall than all other baselines, except for the B-SZZ algorithm.

\find{
{\bf RQ-1:}
\appname is more precise in identifying bug-inducing commits compared to all baselines, with an increase in precision from 2.4\% to 9.2\%. Additionally, \appname achieves a significant enhancement in F1-score, increasing by 6.9\% to 16.0\% compared to the best baselines. It also exhibits more consistent performance across the three datasets. Furthermore, \appname improves both precision and F1-score without a substantial sacrifice in recall.
}

\subsection{RQ2. Effectiveness of key components in \appnamebold}\label{rq2}

\begin{table*}[!t]
\caption{The performance comparisons in ablation study}\label{tab:rq3-result}
\vspace{-3mm}
\resizebox{0.70\textwidth}{!}{

\begin{tabular}[c]{lccccccccc}
\toprule
\multirow{2}{*}{\textbf{Model}} & \multicolumn{3}{c}{\textbf{DS\_LINUX}} & \multicolumn{3}{c}{\textbf{DS\_GITHUB}}  & \multicolumn{3}{c}{\textbf{DS\_APACHE}}   \\ 

& \makecell[l]{Precision} & \makecell[r]{Recall} & \makecell[r]{F1-score} & \makecell[l]{Precision} & \makecell[r]{Recall} & \makecell[r]{F1-score} & \makecell[l]{Precision} & \makecell[r]{Recall} & \makecell[r]{F1-score}  \\

\midrule

\makecell[l]{\appname-raw} & \makecell[c]{0.470} & \makecell[c]{\textbf{0.609}} & \makecell[c]{0.531} & \makecell[c]{0.441} & \makecell[c]{\textbf{0.659}} & \makecell[c]{0.528} & \makecell[c]{0.402} & \makecell[c]{0.393} & \makecell[c]{0.397}  \\
\midrule

\makecell[l]{\appname-r} & \makecell[c]{0.621} & \makecell[c]{0.520} & \makecell[c]{0.566} & \makecell[c]{0.669} & \makecell[c]{0.609} & \makecell[c]{0.637} & \makecell[c]{0.599} & \makecell[c]{0.382} & \makecell[c]{0.467} \\
\midrule

\makecell[l]{\appname-re} & \makecell[c]{0.560} & \makecell[c]{0.511} & \makecell[c]{0.534} & \makecell[c]{0.633} & \makecell[c]{0.567} & \makecell[c]{0.598} & \makecell[c]{0.584} & \makecell[c]{0.383} & \makecell[c]{0.463} \\
\midrule

\makecell[l]{\appname-c} & \makecell[c]{{0.668}} & \makecell[c]{0.450} & \makecell[c]{0.538} & \makecell[c]{\textbf{0.691}} & \makecell[c]{0.498} & \makecell[c]{0.579} & \makecell[c]{\textbf{0.644}} & \makecell[c]{0.316} & \makecell[c]{0.424} \\

\midrule

\makecell[l]{\appname-h} & \makecell[c]{\textbf{0.680}} & \makecell[c]{0.379} & \makecell[c]{0.487} & \makecell[c]{0.564} & \makecell[c]{0.307} & \makecell[c]{0.397} & \makecell[c]{0.523} & \makecell[c]{0.195} & \makecell[c]{0.284} \\

\midrule

\makecell[l]{\appname} & \makecell[c]{0.628} & \makecell[c]{{0.552}} & \makecell[c]{\textbf{0.588}} & \makecell[c]{{0.671}} & \makecell[c]{{0.626}} & \makecell[c]{\textbf{0.647}} & \makecell[c]{0.610} & \makecell[c]{\textbf{0.398}} & \makecell[c]{\textbf{0.482}}  \\

\bottomrule
\end{tabular}

}
\vspace{-3mm}
\end{table*}

\begin{figure}[t]
  \centering
  \includegraphics[width=0.40\linewidth]{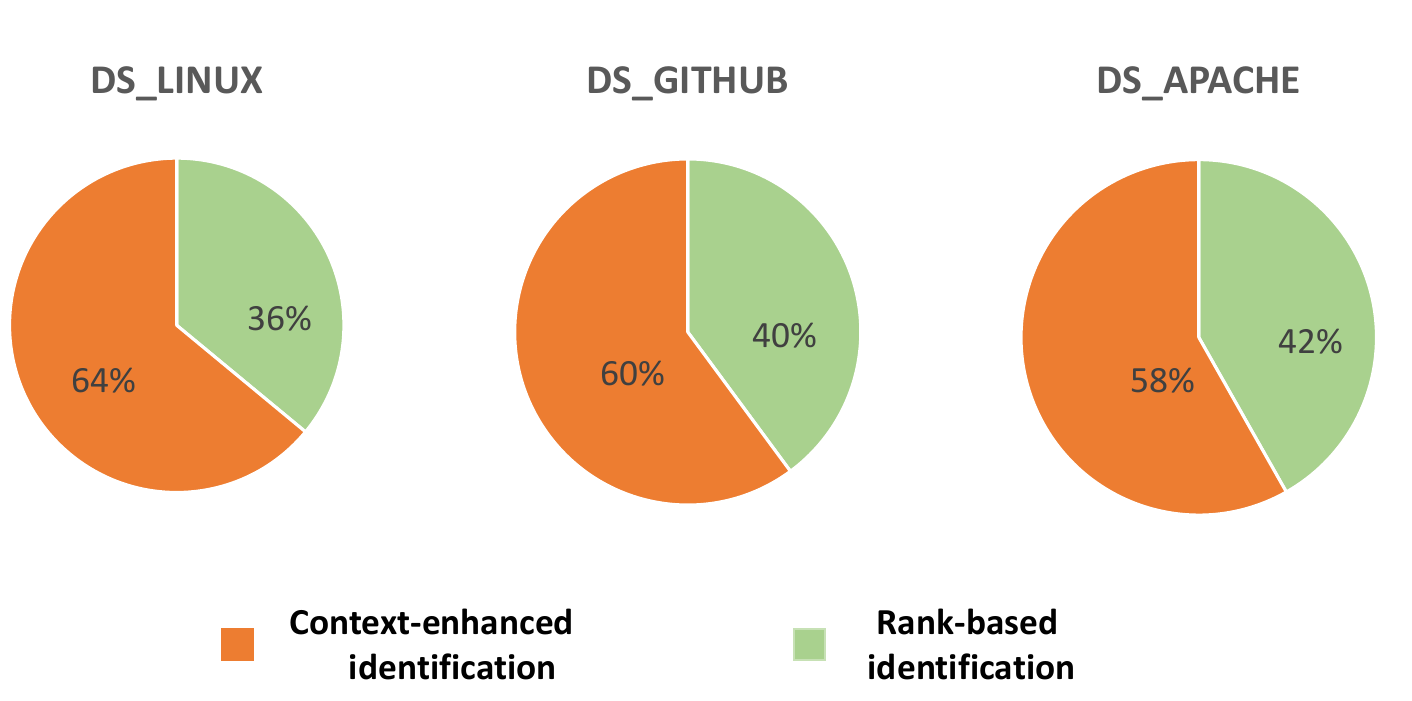}
  \caption{The proportions of the two identification approaches in \appname}\label{fig:strategy-proportions}
  \vspace{-6mm}
\end{figure}

In this section, we investigate the effectiveness of key components in \appname. In \appname-raw, we implement the most basic setting. First, we use the LLM to analyze the root cause of the bug, and then locate the buggy statements based on this root cause. Finally, we trace back the buggy statements to identify the commits that introduced them, marking these as bug-inducing commits.
In \appname-r, we remove context-enhanced assessment and apply rank-based identification across all test cases. The \appname-re variant is built on \appname-r by providing the LLM with expanded context during rank-based identification, rather than the original patch.
In \appname-c, we similarly exclude context-enhanced assessment but utilize context-enhanced identification in all scenarios. The \appname-h variant is based on \appname-c. It omits the hint and provides only the context of the commit along with the root cause of the bug when asking the LLM to determine whether a given commit contains the bug. This allows us to evaluate the contribution of the hint to the LLM's ability to identify the presence of the bug.

Table~\ref{tab:rq3-result} presents the performance of the \appname and its variants in identifying bug-inducing commits. The best results are highlighted in bold. As shown in the table, the \appname-raw variant does not perform very well. This shows that utilizing the LLMs directly on the SZZ algorithm can not improve the performance too much. For example, \appname-raw only improves the F1-socre by 4.7\% compared to the best baseline R-SZZ and B-SZZ in DS\_LINUX and it performs much worse than NeuralSZZ in DS\_APACHE. 
The \appname-r variant, which adopts rank-based identification in all scenarios, outperforms \appname-raw in F1-score across all three datasets, with improvements ranging from 6.5\% to 20.6\%. In contrast, \appname-re performs worse than \appname-r in almost all metrics, indicating that if the LLM cannot comprehend the bug, providing additional context undermines its performance.
The table also shows that \appname-c, which employs context-enhanced identification in all test cases, improves precision compared to \appname-r. However, it may produce empty results in some test cases, leading to lower recall.
Additionally, omitting the hint during context-enhanced identification has a significantly negative impact on  recall, as evidenced by the performance of \appname-h, which is inferior to \appname-c in both recall and F1-score.

Overall, \appname outperforms all other variants in F1-score, highlighting that the combination of rank-based identification and context-enhanced identification can enhance performance. This also demonstrates the effectiveness of the context-enhanced assessment which evaluates the LLM's capabilities and determines the appropriate identification approach.
We also provide the proportions of the two identification approaches in \appname across all three datasets. \begin{coloredtext}As shown in Figure~\ref{fig:strategy-proportions}, context-enhanced identification is used  more frequently in all three datasets. In DS\_LINUX , 64\% of the total test cases utilize context-enhanced identification while in DS\_GITHUB and DS\_APACHE about 60\% of test cases adopt this approach. 
\end{coloredtext}

\find{
{\bf RQ-2:}
The key designs of \appname, including the context-enhanced assessment, the context-enhanced identification and the rank-based identification, all contribute to the overall performance. Compared to the rank-based identification, the context-enhanced identification makes a greater contribution. Furthermore, the hint notably enhances the LLM's ability to determine whether a commit contains the bug.  Additionally, utilizing LLMs directly in the SZZ algorithm does not significantly improve performance.
}

\subsection{RQ3. Effectiveness of \appnamebold on other LLMs} \label{rq3}
In this research question, we aim to examine whether the core ideas of our approach (e.g., preparation, context-enhanced ability check, and commits identification) can be applied to other open-source large language models. For evaluation, we choose two additional open-source LLMs: llama3-8b and llama3-70b. The configurations of these two LLMs are described in Section~\ref{experiment-setting}.

\begin{table*}[!t]\vspace{-2mm}
\caption{The performance comparison between different language models}\label{tab:rq2-result}
\vspace{-3mm}
\resizebox{0.75\textwidth}{!}{

\begin{tabular}[c]{lccccccccc}
\toprule
\multirow{2}{*}{\textbf{Model}} & \multicolumn{3}{c}{\textbf{DS\_LINUX}} & \multicolumn{3}{c}{\textbf{DS\_GITHUB}}  & \multicolumn{3}{c}{\textbf{DS\_APACHE}}   \\ 

& \makecell[l]{Precision} & \makecell[r]{Recall} & \makecell[r]{F1-score} & \makecell[l]{Precision} & \makecell[r]{Recall} & \makecell[r]{F1-score} & \makecell[l]{Precision} & \makecell[r]{Recall} & \makecell[r]{F1-score}  \\

\midrule

\makecell[l]{llama3-8b} & \makecell[c]{0.607} & \makecell[c]{0.536} & \makecell[c]{0.569} & \makecell[c]{0.648} & \makecell[c]{0.595} & \makecell[c]{0.620} & \makecell[c]{0.567} & \makecell[c]{0.371} & \makecell[c]{0.449}  \\
\midrule

\makecell[l]{llama3-70b} & \makecell[c]{\textbf{0.645}} & \makecell[c]{0.551} & \makecell[c]{\textbf{0.594}} & \makecell[c]{0.657} & \makecell[c]{0.611} & \makecell[c]{0.633} & \makecell[c]{\textbf{0.612}} & \makecell[c]{0.397} & \makecell[c]{\textbf{0.482}}  \\

\midrule

\makecell[l]{gpt-4o-mini} & \makecell[c]{0.628} & \makecell[c]{\textbf{0.552}} & \makecell[c]{0.588} & \makecell[c]{\textbf{0.671}} & \makecell[c]{\textbf{0.626}} & \makecell[c]{\textbf{0.648}} & \makecell[c]{0.610} & \makecell[c]{\textbf{0.398}} & \makecell[c]{\textbf{0.482}}  \\

\bottomrule
\end{tabular}

}
\vspace{-3mm}
\end{table*}

Table~\ref{tab:rq2-result} presents the performance of different LLMs across three datasets. Among all LLMs, llama3-8b performs the worst. Despite this, it still outperforms all baselines that are not based on LLMs in RQ1, suggesting that our method can be effectively applied to other large language models. Llama3-70b outperforms llama3-8b in all datasets, which is understandable given that llama3-70b contains more parameters. This also suggests that better LLMs can enhance the performance of \appname.

Compared to gpt4o-mini, llama3-70b performs better in DS\_LINUX, worse in DS\_GITHUB, and similarly in DS\_APACHE. In DS\_LINUX, llama3-70b outperforms gpt-4o-mini in precision and F1-score. To understand the performance differences, we investigate the test cases where the two models produced different results. We find that llama3-70b tends to be conservative when locating buggy statements and produces an empty result if it is uncertain. For instance, llama3-70b produces results for 1,334 test cases in DS\_LINUX, while gpt-4o-mini produces results for 1,373 test cases. In DS\_LINUX, gpt-4o-mini and llama3-70b identified almost the same number of true bug-inducing commits, resulting in higher precision for llama3-70b. In DS\_GITHUB, many bug-inducing commits can only be found by tracing back unmodified code statements, while llama3-70b tends to conservatively identify deleted lines as buggy statements. Therefore, in DS\_GITHUB, gpt-4o-mini outperforms llama3-70b in both precision and recall.

\find{
{\bf RQ-3:}
The core ideas of \appname can be applied to other large language models and better LLMs can enhance the performance of \appname.

}

%% file: tex/discussion.tex
\subsection{Failure Analysis}

\begin{figure}[t]
  \centering
  \includegraphics[width=0.85\linewidth]{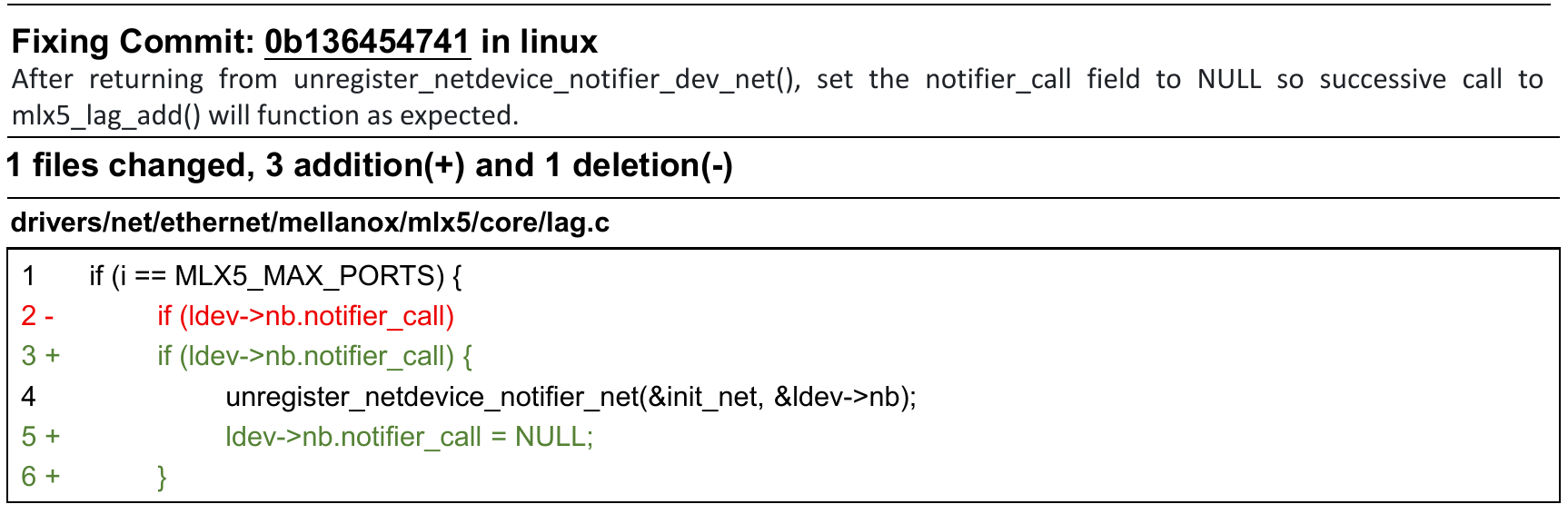}
  \vspace{-3mm}
  \caption{An example where \appname fails to choose the correct bug-inducing commit among candidates.}\label{fig:diss1_1}
  \vspace{-7mm}
\end{figure}

In this section, we manually analyze the test cases where \appname fails to identify their bug-inducing commits correctly. We randomly select 50 test cases from all failed test cases.
After the analysis, we summarize the failed reasons as follows:

\noindent\textbf{Bug-inducing commits cannot be found even by tracing back all lines in the expanded context.} As we mentioned above, \appname provides LLMs with more context to help them locate buggy statements. If changes are within functions, we expand the context by providing LLMs with the full content of the function. Otherwise, we provide it with the three nearest code statements to the changed lines. However, there are still some bug-inducing commits that cannot be found even by tracing back all lines in the expanded context.

In 27 test cases, \appname fails to find bug-inducing commits due to this issue.  In eight of these cases, the bug-fixing commits and their corresponding bug-inducing commits modify completely different files. In the remaining cases, the bug-fixing and bug-inducing commits modify the same files but different functions. Although expanding the context further might help find these bug-inducing commits, an excessively long context will undermine the overall performance of \appname, as mentioned earlier.

\noindent\textbf{Failing to locate buggy statements correctly.} \appname fails to find correct bug-inducing commits in 11 cases due to this issue. One main reason for this is that \appname focuses only on modified lines in the context. Sometimes, it is the unmodified code statements in the context that lead to the bug. Although we provide more context to LLMs to avoid this issue, they still face challenges in accurately locating buggy statements.

\noindent\textbf{Failing to choose the correct bug-inducing commit among all candidates.} The remaining 12 test cases fail because \appname does not identify the correct bug-inducing commit among all candidates. The main reason is that \appname fails to determine whether a commit contains the bug. The bug-fixing commit $0b136454741$ in Figure~\ref{fig:diss1_1} is a typical example.  From the commit message, \appname first locates two buggy statements, lines 2 and 4, resulting in two candidate commits: $e387f7d5fcc$ and $7907f23adc1$. Commit $e387f7d5fcc$ introduces line 4 and replaces the former function \code{unregister\_netdevice\_notifier\_dev\_net} with the current function \code{unregister\_netdevice\_notifier\_net} . The LLM determines that although returning from the function \code{unregister\_netdevice\_notifier\_dev\_net} requires setting the \code{notifier\_cal} field to null, this does not imply the same for the function \code{unregister\_netdevice\_notifier\_net} as they are different functions. Therefore, it determines that commit $e387f7d5fcc$\texttt{\^}1 does not contain the bug, and \appname incorrectly identifies $e387f7d5fcc$ as the bug-inducing commit.

\subsection{Data Leakage}

\begin{table*}[!t]\vspace{3mm}
\caption{The performance comparisions between all methods in extended datasets}\label{tab:new-data-result}
\vspace{-3mm}
\resizebox{0.50\textwidth}{!}{

\begin{tabular}[c]{lccccccccc}
\toprule
\multirow{2}{*}{\textbf{Method}} & \multicolumn{3}{c}{\textbf{DS\_LINUX}}  & \multicolumn{3}{c}{\textbf{DS\_GITHUB}}   \\

& \makecell[l]{Precision} & \makecell[r]{Recall} & \makecell[r]{F1-score} & \makecell[l]{Precision} & \makecell[r]{Recall} & \makecell[r]{F1-score}  \\
\midrule

\makecell[l]{B-SZZ} & \makecell[c]{0.443} & \makecell[c]{\textbf{0.592}} & \makecell[c]{0.507} & \makecell[c]{0.416} & \makecell[c]{\textbf{0.669}} & \makecell[c]{0.513}  \\
\midrule

\makecell[l]{AG-SZZ} & \makecell[c]{0.480} & \makecell[c]{0.532} & \makecell[c]{0.505} & \makecell[c]{0.480} & \makecell[c]{0.581} & \makecell[c]{0.526} \\
\midrule

\makecell[l]{MA-SZZ} & \makecell[c]{0.407} & \makecell[c]{0.532} & \makecell[c]{0.461} & \makecell[c]{0.401} & \makecell[c]{0.595} & \makecell[c]{0.479}  \\
\midrule

\makecell[l]{R-SZZ} & \makecell[c]{0.601} & \makecell[c]{0.458} & \makecell[c]{{0.519}} & \makecell[c]{0.621} & \makecell[c]{0.520} & \makecell[c]{{0.566}}  \\
\midrule

\makecell[l]{L-SZZ} & \makecell[c]{0.564} & \makecell[c]{0.430} & \makecell[c]{0.488} & \makecell[c]{0.548} & \makecell[c]{0.459} & \makecell[c]{0.500} \\
\midrule

\makecell[l]{\appname} & \makecell[c]{\textbf{0.642}} & \makecell[c]{0.579} & \makecell[c]{\textbf{0.609}} & \makecell[c]{\textbf{0.646}} & \makecell[c]{0.604} & \makecell[c]{\textbf{0.624}}  \\

\bottomrule
\end{tabular}

}
\vspace{-5mm}
\end{table*}
\begin{coloredtext}
    
Since all three datasets were collected before the release of LLMs like GPT-4o-mini~\cite{gpt-4o-mini}, there is a potential issue that the performance improvement of \appname may result from data leakage. To address this concern, we extend two of the datasets.
To extend DS\_GITHUB, we follow the same approach as the original paper~\cite{rosa2021evaluating}. Specifically, we iterate through all commits in the dataset’s projects, identifying those whose commit messages contain keywords such as "fix" "bug" and "introduce". These commits are added to a candidate list. As GPT-4o-mini’s knowledge is limited to data available up to October 2023~\cite{gpt-4o-mini}, we exclude all commits dated before this cutoff. We initially identify 186 candidate commits. Each commit is then manually reviewed to confirm its relevance to bug fixing and to ensure the existence of corresponding bug-inducing commits. After this filtering process, we obtain 148 verified bug-fixing commits and their associated bug-inducing commits.
To extend the DS\_LINUX dataset, we adopt the method used in the original study~\cite{lyu2024evaluating}. Bug-fixing commits in Linux typically include the keyword "Fixes:" followed by the commit ID of the bug-inducing commit. Using a regular expression, we identify commits with this pattern. Commits dated before October 2023 are excluded, resulting in 9,913 bug-fixing commits. From these, we randomly sample 500 commits for the experiment, ensuring a 95\% confidence level and a margin of error under 5\%, the same as section~\ref{experiment-setting}.
For DS\_APACHE, extending the dataset is more challenging as it is based on bug reports that lack a fixed format for identifying bug-inducing commits. In the original study, all bug reports were manually analyzed, which required significant human effort. Given that our objective is to demonstrate that the performance improvement of \appname is not due to data leakage, the extended datasets of DS\_GITHUB and DS\_LINUX are sufficient for this purpose. Therefore, we opt not to extend DS\_APACHE.

Table~\ref{tab:new-data-result} presents the experimental results of \appname and the baselines for identifying bug-inducing commits across the extended datasets. The best results are highlighted in bold. Among the baselines, the R-SZZ algorithm achieves the best performance in two datasets. In DS\_LINUX, R-SZZ performs comparably to the B-SZZ and AG-SZZ algorithms, while in DS\_GITHUB, it significantly outperforms all other baselines.
From the table, we observe that \appname consistently outperforms all baselines. Specifically, in DS\_LINUX, it improves precision by 6.8\% and the F1-score by 17.3\% compared to the best-performing baseline. Similarly, in DS\_GITHUB, \appname improves precision by 4.0\% and the F1-score by 10.2\%. These results demonstrate that \appname's performance improvements are not caused by data leakage.

\end{coloredtext}

\subsection{Scalability}

\begin{table*}[!t]\vspace{-3mm}
\caption{Statistics related to the scalability of \appname}\label{tab:scalability}
\vspace{-3mm}
\resizebox{0.40\textwidth}{!}{

\begin{tabular}[c]{lccccccccc}
\toprule
\textbf{DATASET}  & \textbf{llm calls} & \textbf{token numbers} & \textbf{time}\\

\midrule

\makecell[l]{DS\_LINUX}  & \makecell[c]{9.8} & \makecell[c]{14,489}  & \makecell[c]{30.43s} \\
\midrule

\makecell[l]{DS\_GITHUB}  & \makecell[c]{10.3} & \makecell[c]{14,890} & \makecell[c]{20.20s}  \\

\midrule

\makecell[l]{DS\_APACHE}  & \makecell[c]{13.67} & \makecell[c]{24,023} & \makecell[c]{28.14s}  \\

\bottomrule
\end{tabular}

}
\vspace{-4mm}
\end{table*}

\begin{coloredtext}
Table~\ref{tab:scalability} presents statistics on the scalability of \appname, including the average number of LLM calls, the average number of tokens consumed, and the average time required to process a bug-fixing commit. The average number of LLM calls and the average token consumption are primarily determined by the size of the bug-fixing commit. Bug-fixing commits in DS\_APACHE are generally larger than those in DS\_GITHUB and DS\_LINUX, which results in higher LLM call frequencies and greater token consumption for DS\_APACHE.

Additionally, the table shows that \appname requires approximately 30 seconds to handle a bug-fixing commit. This is longer than the processing time of some baselines, such as the B-SZZ algorithm, because these baselines rely on basic assumptions or simple heuristic rules. However, when compared to more complex techniques like RA-SZZ, which employs program analysis to detect refactorings, \appname is significantly faster. Our experiment indicates that RA-SZZ takes an average of 78.4 seconds to process a single bug-fixing commit, nearly 2.6 times longer than \appname.

The total time cost of \appname consists of two components: the time for LLM calls and the time required to retrieve relevant project information. For example, \appname retrieves file contents from specific commits in the project, which contributes to the time cost. This explains why the average processing time for bug-fixing commits in DS\_LINUX is longer than that in DS\_GITHUB and DS\_APACHE, even though DS\_LINUX requires fewer LLM calls. The Linux project’s large size increases the time required for information retrieval, leading to higher overall processing times. Although this larger size results in increased time, the overal time remains acceptable, indicating that \appname is feasible for large projects.  We also investigate whether there are any extremely long bug-fixing commits that exceed the LLM's token limit. Our results show that there is only one such commit in the Hadoop project, and it has minimal impact on overall performance.

\end{coloredtext}

\begin{coloredtext}
\subsection{Bugs that LLMs fail to understand}
In this section, we analyze the test cases in which LLMs fail to understand the bug and fall back to rank-based identification. We randomly select 50 such test cases from the total of 797 failed cases and manually examine the characteristics of the bugs. Among these, 38 test cases are from DS\_LINUX, 7 from DS\_GITHUB, and 5 from DS\_APACHE. We categorize the bugs into two types: those with an excessively large context and those requiring subtle changes to resolve.

\noindent\textbf{Bugs involving an excessively large context.} \appname performs context refinement before requiring the LLM to determine whether the commit contains a bug. However, even after refinement, the context can remain excessively large. In 32 of the analyzed test cases, the context ranges from 330 to 1,887 lines of code. This overwhelming long context hinders the LLM's ability to accurately detect the bug. 

\noindent\textbf{Fixing the bug requires subtle changes.} The remaining 18 failed test cases arise from the LLM's difficulty in detecting subtle changes. For example, some bug-fixing commits only reorder code statements. In such cases, the LLM may incorrectly assume that both versions contain the same statements, misidentifying them as buggy. Similarly, bug fixes involving minimal changes, such as altering a single word in a long string, are also challenging for the LLM to detect.
\end{coloredtext}

\subsection{Can \appnamebold find extra bug-inducing commits?}

\begin{table*}[!t]
\caption{The statistics of extra bug-inducing commits identified by \appname}\label{tab:diss2}
\vspace{-3mm}
\resizebox{0.45\textwidth}{!}{

\begin{tabular}[c]{lccccccccc}
\toprule
\textbf{Count} & \textbf{DS\_LINUX} & \textbf{DS\_GITHUB} & \textbf{DS\_APACHE}  \\

\midrule

\makecell[l]{only-additions} & \makecell[c]{97} & \makecell[c]{15} & \makecell[c]{3}   \\
\midrule

\makecell[l]{with-deletions} & \makecell[c]{25} & \makecell[c]{12} & \makecell[c]{6}   \\

\midrule

\makecell[l]{total} & \makecell[c]{122} & \makecell[c]{27} & \makecell[c]{9}   \\

\bottomrule
\end{tabular}

}
\vspace{-5mm}
\end{table*}

In this section, we examine whether \appname can identify extra commits that all baselines cannot. Table~\ref{tab:diss2}
presents the statistics of the extra bug-inducing commits that baselines fail to find. \appname identifies 109 extra bug-inducing commits in DS\_LINUX, 31 in DS\_GITHUB, and 10 in DS\_APACHE, accounting for 7.8\%, 7.4\%, and 2.5\% of the total bug-inducing commits, respectively.

We classify these bug-fixing commits into two categories: those containing only added lines and those containing deleted lines. From the table, we observe that \appname effectively identifies bug-inducing commits from those with only added lines. Additionally, techniques employed in \appname, such as context expanding, facilitate the discovery of extra bug-inducing commits from bug-fixing commits with deleted lines. In DS\_LINUX, the majority of extra bug-inducing commits are identified from bug-fixing commits with only added lines. Conversely, in DS\_GITHUB, the number of extra bug-inducing commits found from bug-fixing commits with only added lines is nearly equal to those identified from bug-fixing commits with deleted lines.

\subsection{Threats to Validity}

\noindent\textbf{Internal Validity.} The LLMs may produce random outputs during the experiment. To minimize bias, we set the same parameters for all models. Additionally, we repeat the experiment three times and select the majority result as the final output. We also conduct our experiment on large-scale datasets to counteract the randomness. These datasets consist of two programming languages and a total of 2,104 test cases.

\noindent\textbf{External Validity.} One potential limitation is that we implement and evaluate \appname on only two programming languages, C and Java. However, the majority of bug-fixing commits in the three datasets are written in these two languages. Another concern is that DS\_LINUX is created by randomly selecting test cases from the original dataset. To mitigate bias, we selected a total of 1,500 test cases, achieving over a 95\% confidence level with a confidence interval of 3. \begin{coloredtext}
The third threat is the assumption that most bugs are fully fixed and introduced with a single commit. While this assumption holds for most test cases in the three datasets, it may not reflect all real-world scenarios. In the future, we plan to collect additional datasets to address this limitation.\end{coloredtext}

%% file: tex/related_work.tex
\noindent\textbf{LLMs in SE.} LLMs have been applied to numerous tasks in software engineering~\cite{hou2023large,fan2023large}, such as code generation, software testing and software mountainance. In code generation, researchers have proposed generation models like CodeX~\cite{chen2021evaluating}, AlphaCode~\cite{li2022competition} and Codegen~\cite{nijkamp2022codegen}. They have also improved the performance of code generation using techniques such as  chain of thought reasoning~\cite{jiang2023selfevolve,zhang2023self}, static analysis~\cite{ahmed2024automatic} and finetuning~\cite{shin2023prompt}. LLMs can also be used to generate new test cases, showing higher coverage~\cite{hu2023augmenting,deng2023large}. Combing with techniques such as differential testing,  they can  generate more failure-inducing test cases~\cite{li2023nuances}. In software maintenance, LLMs can be used in tasks such as fault localization~\cite{kang2023preliminary,wu2023large}, bug reproducing~\cite{feng2024prompting}, bug severity predicting~\cite{mashhadi2023method} and program repair~\cite{xia2022less,xia2023conversational,xia2023keep}.

\noindent\textbf{SZZ algorithm evaluation.} The SZZ algorithm and its variants have been extensively evaluated by many researchers. Initially, evaluations are based on datasets manually annotated by researchers~\cite{davies2014comparing}. However, building such datasets is time-consuming and may not yield accurate results. To address these challenges, researchers have proposed datasets based on developers' annotations. They extract these annotations from bug reports~\cite{wen2019exploring} and commit messages~\cite{rosa2021evaluating,lyu2024evaluating}.

\noindent\textbf{SZZ algorithm application.} The SZZ algorithms have been widely used in empirical studies, including software quality~\cite{ccaglayan2016effect}, code smells~\cite{palomba2018diffuseness}, code reviews~\cite{kononenko2015investigating,bavota2015four}, and developer collaboration~\cite{bernardi2018relation}. The SZZ algorithm has also been applied to just-in-time defect detection~\cite{kamei2012large,jiang2013personalized,mcintosh2018fix}. Researchers use the SZZ algorithm to identify bug-inducing commits in projects, which are then used to train models and evaluate their effectiveness.

%% file: tex/conclusion_future_work.tex
In this study, we propose a novel approach named \appname, which utilizes large language models (LLMs) to locate bug-inducing commits based on bug-fixing commits. The core idea of \appname is to adopt different approaches for identifying bug-inducing commits based on the LLM's ability to comprehend the bug. During the ability assessment, we provide the LLM with both expanded and refined contexts to assist it in locating buggy statements and determining whether the bug exists. Based on the assessment results, we then employ either rank-based identification or context-enhanced identification. We evaluate \appname using three high-quality datasets, and experimental results show that it outperforms all other baselines in F1-score and can identify extra bug-inducing commits that the baselines cannot detect. In the future, we plan to extend \appname to support additional programming languages and collect more high-quality datasets of bug-fixing commits along with their corresponding bug-inducing commits. Additionally, we intend to fine-tune the LLMs to enhance their ability to comprehend bugs and determine their presence in a commit. Furthermore, we aim to leverage the LLMs to identify bug-inducing commits based on bug-fixing commits, even in cases where they do not modify the same files or functions.